\documentclass[aps,twocolumn,prl,tightenlines,floatfix]{revtex4-1}
\usepackage{hyperref}
\usepackage{graphicx}
\usepackage[english]{babel}
\usepackage{amsmath}
\usepackage{amsfonts}
\usepackage{amssymb}
\usepackage{times}

\begin{document}
%\setpagewiselinenumbers
%\modulolinenumbers[5]
%\linenumbers
%\setlength\columnsep{22pt}

\title{Instability of Fulde-Ferrell-Larkin-Ovchinnikov states in three
  and two dimensions}

\author{Jibiao Wang, Yanming Che, Leifeng Zhang }
\affiliation{Department of Physics and Zhejiang Institute of Modern
  Physics, Zhejiang University, Hangzhou, Zhejiang 310027, China}
\affiliation{Synergetic Innovation Center of Quantum Information and
  Quantum Physics, Hefei, Anhui 230026, China} 
\author{Qijin Chen}
\email[Corresponding author: ]{qchen@zju.edu.cn}
\affiliation{Department of Physics and Zhejiang Institute of Modern
  Physics, Zhejiang University, Hangzhou, Zhejiang 310027, China}
\affiliation{Synergetic Innovation Center of Quantum Information and
  Quantum Physics, Hefei, Anhui 230026, China}

\date{\today}

\begin{abstract}
  The exotic Fulde-Ferrell-Larkin-Ovchinnikov (FFLO) states have been
  actively searched for experimentally since the mean-field based FFLO
  theories were put forward half a century ago.  Here we investigate
  the stability of FFLO states against unavoidable pairing
  fluctuations, and conclude that FFLO superfluids cannot exist due to
  their intrinsic instability in three and two dimensions. This
  explains their absence in experimental observations in both
  condensed matter systems and the most recent, more promising
  ultracold atomic Fermi gases with a population imbalance.
\end{abstract}

%\pacs{03.75.Ss,03.75.Hh,67.85.-d,74.25.Dw}

\maketitle

The exotic Fulde-Ferrell-Larkin-Ovchinnikov (FFLO) states , which were
first predicted by Fulde and Ferrell \cite{FF} (FF) and Larkin and
Ovchinnikov \cite{LO} (LO) in an $s$-wave superconductor in the
presence of a Zeeman field over fifty years ago, have attracted
enormous attention in condensed matter physics \cite{LOFF_Review},
including heavy-fermion \cite{Radovan,Kenzelmann}, organic
\cite{ShimaharaJPSJ71,*LebedPRB82} and high $T_c$ superconductors
\cite{GrafPRB72,*TingPRL96,*SimonsPRL102,*ChoPRA83,*Ptok2013}, nuclear
matter \cite{SedrakianPRC67} and color superconductivity
\cite{Alford,*Bowers}, and, more recently, in ultracold Fermi gases
\cite{SedrakianPRA72,*Liao,*Sheehy_RPP}. 
Conventional BCS superfluidity originates from Bose-Einstein
condensation (BEC) of Cooper pairs at zero momentum. In contrast, in
these exotic states, Cooper pairs condense either at a finite momentum
$\textbf{q}$, with an order parameter of the form of a plane-wave
$\Delta(\textbf{r})=\Delta_{0}e^{i\textbf{q}\cdot\textbf{r}}$ or at
momenta $\pm \textbf{q}$, with an order parameter of the form of a
standing wave
$\Delta(\textbf{r})=\Delta_{0}\cos(\textbf{q}\centerdot\textbf{r})$
for the FF and LO states, respectively. 

These exotic superfluids have been actively searched for over the past
half century. 
In condensed matter systems, the strongest signatures of FFLO states
come from heavy fermion UPd$_2$Al$_3$ \cite{Gloos}, CeRu$_2$
\cite{Huxley} and CeCoIn$_5$ \cite{Radovan,Kenzelmann}. However,
Refs. \cite{Gloos} and \cite{Huxley} were shown to be inconsistent
with theory \cite{NormanFFLOcomment,Tenya}. Radovan \textit{et al.}
\cite{Radovan} assumed an incorrect FFLO wavevector direction
perpendicular to the highly two-dimensional Fermi surface, and their
claim seemed also to have been dismissed recently by Kenzelmann
\textit{et al.}  \cite{Kenzelmann}, who also noticed discrepancies
between theory and their own experimental observations. Thus far,
\emph{there has been no solid experimental evidence for the FFLO states
from condensed matter systems}.

With the easy tunability of various control parameters, including
interaction, dimensionality, population imbalance as well as mass
imbalance \cite{Review,Bloch_RMP}, ultracold Fermi gases have provided
a much greater opportunity and given rise to a high expectation for
finding the FFLO states.
Despite many theoretical studies in this regard, both in a 3D
homogeneous case
\cite{LOFF1,SR06,huhui06pra,helianyi06prb,CombescotPRB7114,*Yip07pra,FFLO_MF_us}
and in a trap \cite{Machida2,zw07pra,Kinnunen}, the experimental
search for these exotic states in atomic Fermi gases has not been
successful \cite{ZSSK06,Rice1}.
There have also been theoretical studies of FFLO states in more
complex systems, such as Fermi-Fermi mixtures
\cite{Stoof09prl,*Stoof10pra,Stoof12pra,FFLO_MF_us}
% or equal-mass Fermi gases with spin-orbit coupling
% \cite{chuanwei12,*xiaji13,Pu2013,*Iskin2013,*Yi2013PRL}
or optical lattices
\cite{WuCJPRA83,*Buchleitner2012,*Mendoza2013,*Koga2012,*Torma2012,*ChenAHai2012}.
% Our recent work \cite{FFLO_MF_us} points out that a large mass ratio
% in a Fermi-Fermi mixture may indeed enhance FFLO type of pairing.
%
However, experimentally, the superfluid regime in these complex
systems has yet to be accessed.

%In this paper, we are going to ask and answer why the FFLO states have
%not been observed experimentally.
%%
%While this failure has been largely attributed by many to the
%\emph{small} areas and the \emph{very low} temperature needed for the
%FFLO superfluid phase in the phase diagrams
%\cite{LOFF1,SR06,huhui06pra,zw07pra}, here we conclude that there is a
%deeper reason behind this failure. In particular, the FFLO states are
%intrinsically unstable in 3D and 2D.
%%
%Indeed, both the FF and LO states as well as higher order crystalline
%states in the literature are essentially constructed at the mean field
%level, and their stability has never been properly tested against
%pairing fluctuations.

In this paper, we will reveal the deep reason why the FFLO states have
not been observed experimentally. 
Here we investigate the stability of FFLO states against ubiquitous
pairing fluctuations, first with simple arguments based on general
physical grounds, and then using a concrete pairing fluctuation theory
\cite{Chen2,Review}, which has been applied successfully to the
BCS-BEC crossover physics.  We find that FFLO states are indeed
intrinsically unstable at any finite temperature $T$ due to pairing
fluctuations. This conclusion can be drawn using rival approaches of
pairing fluctuation theories as well. We shall mainly work with the 3D
case and readily generalize to 2D.
We note that both the FF and LO states as well as higher order crystalline
states in the literature are essentially constructed at the mean field
level, and their stability has never been \emph{properly} tested against
pairing fluctuations.

Now consider a single minority fermion in the presence of a majority
Fermi sphere in homogeneous 3D continuum, assuming an equal mass for
both majority and minority fermions. When the pairing strength is
weak, the ground state is a polaron in the Fermi sea.  When the
interaction becomes just strong enough, the minority atom will pair
with a majority atom near the Fermi surface to form a (meta-)stable
pair. To minimize the system energy, the pair dispersion will reach a
minimum at a finite momentum of $q\approx k_F$, where $k_F$ is the
majority Fermi wave vector \cite{footnoteonBECpairing}. Similar things
will happen for a two component gas with a high population
imbalance. For weak interactions, the ground state will be minority
polarons moving in the majority Fermi sea. For very strong
interactions in the BEC regime, a polarized Sarma superfluid will
emerge at low $T$. For intermediate pairing strengths, where the
majority Fermi surface still exists, (meta-)stable Cooper pairs will
form at finite $T$, with a dispersion minimizing at a finite momentum
$q$. These pairs will first form in the normal state without phase
coherence, moving in all possible directions. As $T$ decreases, the
system will either phase separate into a 50-50 mixture forming a BCS
superfluid plus a majority normal Fermi gas, or try to enter an FF or
LO state. Now that the pair dispersion minimizes at a finite momentum,
i.e., on a 2D spherical surface $\mathbb{S}^2$ in the momentum space,
one finds immediately that no condensation is needed at any finite
$T$, in order to satisfy the pair density constraint. Alternatively,
the pairing fluctuations will destroy any tendency of Bose
condensation of the pairs.

The key here is that in the momentum space, FFLO states can be
regarded as condensation of pairs (at finite momenta) whose energy
minimizes on a 2D sphere $\mathbb{S}^2$. Such a 2D Bose surface for
the pair dispersion has a finite density of states (DOS), leading to
an effective reduction of the dimensionality for the pairs from 3D to
2D, so that these fluctuations will destroy any attempt for
condensation, in accord with the Mermin-Wagner theorem. This argument
can be readily extended to the case of an optical lattice, where the
2D sphere is to be replaced by a 2D constant-energy surface.

Next we examine what would happen if one forces a symmetry breaking
into an FFLO state at low $T$ with a wavevector $\mathbf{q}$ pointing
in the symmetry breaking direction, as in a mean-field treatment.  To
this end, we proceed with a concrete theoretical formalism of pairing
fluctuations in two-component homogeneous Fermi gases with a
population imbalance in 3D continuum. We begin by presenting the
mean-field solutions, and then show that the mean-field FFLO phase
will eventually be destroyed by pairing fluctuations.
We shall first restrict ourselves to regular symmetry breaking, i.e.,
with only one wavevector $\mathbf{q}$, which corresponds to the FF
states.
As usual, we work a short-range contact potential of strength $U<0$.
In this assumed FF phase,
momentum $\textbf{k}$ pairs with $\textbf{q}-\textbf{k}$ and thus the
condensed Cooper pairs have a nonzero center-of-mass momentum
$\textbf{q}$. Note that setting $\textbf{q}=0$ would give us the
formalism for the Sarma superfluid state. The dispersion of free atoms
is given by $\xi_{\textbf{k},\sigma}=
\mathbf{k}^{2}/2m_{\sigma}-\mu_{\sigma}$, where $m_{\sigma}$ and
$\mu_{\sigma}$ are the mass and chemical potential for (pseudo)spin
$\sigma=\uparrow,\downarrow$, respectively.  We set the volume $V=1$,
$\hbar=k_{B}=1$.

In order to self-consistently treat the pairing fluctuation effects,
we use a pairing fluctuation theory previously developed \cite{Chen2}
for treating the pseudogap phenomena in high $T_c$ superconductors,
which has later been extended successfully to address a variety of
ultracold Fermi gas experiments without
\cite{Review,ReviewLTP-Full,OurRFReview} and with population
\cite{Chien06,chen07prb} and/or mass imbalances
\cite{Guo2009PRA,wang13pra}.  
Within this theory, the BCS mean-field solution of the FF states can
be obtained from the combined gap equation and number equations by
neglecting the pseudogap equation.  In addition, in the superfluid
phase, the effective chemical potential of the pairs $\mu_{pair}$
vanishes, which guarantees that the pair excitation energy is
\emph{not} gapped in the superfluid phase.

Now we shall present our self-consistent equations for the mean-field
solution from this theory. 
Since the pair dispersion minimizes at $\mathbf{q}\ne 0$ for the FF
states, the Thouless criterion for pairing instability now reads
$t^{-1}_{pg}(0,\textbf{q})=U^{-1}+\chi(0,\textbf{q})=0$, with
$t_{pg}(P)$ being the $T$-matrix,
$\chi(P)=\sum_{K,\sigma}G_{0\sigma}(P-K)G_{\bar{\sigma}}(K)/2$ the
pair susceptibility, $G_0(K)$ and $G(K)$ the bare and full Green's
functions, respectively, and
$G^{-1}_{0\sigma}(K)=i\omega_{n}-\xi_{\textbf{k},\sigma}$. (Here spin
${\bar{\sigma}}$ is the opposite of spin $\sigma$). \emph{We refer the
  readers to Ref.~\cite{wang13pra} for the convention on notations.}
The self-energy \cite{Review} takes approximately the simple BCS-like
form, $\Sigma_{\sigma}(K)=-\Delta^{2}G_{0\bar{\sigma}}(Q-K)$, with $
Q\equiv (0, \mathbf{q})$.
Therefore, we have
\begin{subequations}
\label{eq:G}
\begin{eqnarray}
G_{\uparrow}(K)&=&\frac{u_{\textbf{k}}^{2}}{i\omega_{n}-E_{\textbf{k},\uparrow}}+
\frac{v_{\textbf{k}}^{2}}{i\omega_{n}+E_{\textbf{k},\downarrow}},\\
G_{\downarrow}(K)&=&\frac{u^{2}_{\mathbf{q-k}}}{i\omega_{n}-E_{\mathbf{q-k},\downarrow}}+
\frac{v^{2}_{\mathbf{q-k}}}{i\omega_{n}+E_{\mathbf{q-k},\uparrow}},
\end{eqnarray}
\end{subequations}
where $u_{\textbf{k}}^{2}=(1+\xi_{\textbf{kq}}/E_{\textbf{kq}})/2$,
$v_{\textbf{k}}^{2}=(1-\xi_{\textbf{kq}}/E_{\textbf{kq}})/2$,
$E_{\textbf{kq}}=\sqrt{\xi_{\textbf{kq}}^{2}+\Delta^{2}}$, and
$E_{\textbf{k},\uparrow}=E_{\textbf{kq}}+\zeta_{\textbf{kq}}$,
$E_{\textbf{k},\downarrow}=E_{\textbf{kq}}-\zeta_{\textbf{kq}}$,
$\xi_{\textbf{kq}}=(\xi_{\textbf{k},\uparrow}
+\xi_{\textbf{q}-\textbf{k},\downarrow})/2$,
$\zeta_{\textbf{kq}}=(\xi_{\textbf{k},\uparrow}
-\xi_{\textbf{q}-\textbf{k},\downarrow})/2$.  Here the quasiparticle
dispersion $E_{\mathbf{k},\sigma}$ may not be gapped. In the presence
of mass imbalance, $u_{\mathbf{k}}^2 \ne u_{\mathbf{q-k}}^2$, unlike
the equal-mass case.
With the BCS form for the Green's functions Eq.~(\ref{eq:G}), the
Thouless criterion becomes
\begin{equation}
  \frac{m_{r}}{2\pi a}=\sum_{\textbf{k}}\Big[\frac{1}{2\epsilon_{\textbf{k}}^{}}-
  \frac{1-2\bar{f}(E_{\textbf{kq}})}{2E_{\textbf{kq}}}\Big]\,,
  \label{eq:gap}
\end{equation}
where $\epsilon_{\textbf{k}}^{}=k^{2}/4m_{r}$ with reduced mass $m_{r}$. Here
$\bar{f}(x)=[f(x+\zeta_{\textbf{kq}})+f(x-\zeta_{\textbf{kq}})]/2$,
and $f(x)$ is the Fermi distribution function.
Note that $U$ has been replaced by the $s$-wave scattering length $a$
via $U^{-1}=m_{r}/2\pi a-\sum_{\textbf{k}}1/2\epsilon_{\textbf{k}}^{}$.

From the number constraint $n_{\sigma}=\sum_{K}G_{\sigma}(K)$, we can
get the number density $n=n_{\uparrow}+n_{\downarrow}$ and density
difference $\delta n\equiv n_{\uparrow}-n_{\downarrow}$,
\begin{eqnarray}
  n&=&\sum_{\textbf{k}}\Big[\Big(1-\frac{\xi_{\textbf{kq}}}{E_{\textbf{kq}}}\Big)+
  2\bar{f}(E_{\textbf{kq}})\frac{\xi_{\textbf{kq}}}{E_{\textbf{kq}}}\Big], 
  \label{eq:LOFF_neqa}\\
  \delta n&=&\sum_{\textbf{k}}
  \Big[f(E_{\textbf{k},\uparrow})-f(E_{\textbf{k},\downarrow})\Big].
  \label{eq:LOFF_neqb}
\end{eqnarray}
The population imbalance is defined as $\eta = \delta n/n$.

The FFLO wavevector $\mathbf{q}$ can be determined via $\frac{\partial
  \chi(0,\mathbf{p})}{\partial \mathbf{p}}|_\mathbf{p=q}=0$, which is
equivalent to minimizing the thermodynamic potential
$\Omega_{S}$ with respect to $\mathbf{q}$ \cite{FFLO_MF_us}. Then we have 
\begin{eqnarray}
\sum_{\mathbf{k}}\left[\frac{\mathbf{k}}
  {m_{\uparrow}}(n_{\textbf{kq}}+\delta n_{\textbf{kq}})+
  \frac{\mathbf{q-k}}{m_{\downarrow}}
  ( n_{\textbf{kq}}-\delta n_{\textbf{kq}})\right]=0\,,\quad
  \label{eq:min}
\end{eqnarray}
where $n_{\textbf{kq}}$ and $\delta n_{\textbf{kq}}$ are given by the
summands of Eqs.~(\ref{eq:LOFF_neqa}) and (\ref{eq:LOFF_neqb}),
respectively. 
% To address the stability of the mean-field solutions against pairing
% fluctuations, we need to obtain the energy spectrum of the pairs.

Equations (\ref{eq:gap})-(\ref{eq:min}) form a closed set, and can be
used to solve for the mean-field solution of the one-plane-wave FFLO
state, e.g., for ($\mu_\uparrow$, $\mu_\downarrow$, $T_c$,
$\mathbf{q}$) with $\Delta=0$, and for ($\mu_\uparrow$,
$\mu_\downarrow$, $\Delta$, $\mathbf{q}$) at $T<T_c$.  Dropping
Eq.~(\ref{eq:min}) and setting $\mathbf{q}=0$ would lead to mean-field
equations for homogeneous Sarma phases.

At the mean-field level, the FFLO solutions may be further restricted
by the stability condition against phase separation (PS)
\cite{PWY05,Stability,LOFF1},
\begin{equation}
  \frac{\partial^{2}\Omega_{S}}{\partial\Delta^{2}}
  \frac{\partial^{2}\Omega_{S}}{\partial\textbf{q}^{2}}-
  \Big(\frac{\partial^{2}\Omega_{S}}
  {\partial\Delta\partial\textbf{q}}\Big)^{2}>0\,.
\label{eq:sta}
\end{equation}

With the mean-field solutions, one can extract the pair dispersion
${\Omega}_\mathbf{p}$ via a Taylor expansion of the inverse $T$-matrix
\cite{Review}, i.e., $t_{pg}^{-1}(\Omega,\textbf{p})\approx
a_{0}(\Omega-\Omega_\mathbf{p}+\mu_{pair})=0$, after analytic
continuation,
where $\Omega_\mathbf{p}= -[\chi(0,\mathbf{p})-\chi(0,\mathbf{q})]/a_0
\approx B_\parallel({p}_\parallel-{q})^{2}+B_\perp p_\perp^2$ near
$\mathbf{p=q}$.
The coefficients $a_{0}$, $B_\parallel$, $B_\perp$, and $\mu_{pair}$
can be readily derived during the expansion, with $\mu_{pair}=0$ at
$T\leq T_{c}$, and $ \mu_{pair}<0$ above $T_c$. Here the subscripts
``$\parallel$'' and ``$\perp$'' denote parallel with and perpendicular
to the $\mathbf{q}$ direction, respectively.

\begin{figure}
%  \centerline{\includegraphics[clip,width=3.4in] {FF-Unitary_nphys.eps}}
  \centerline{\includegraphics[clip,width=3.4in] {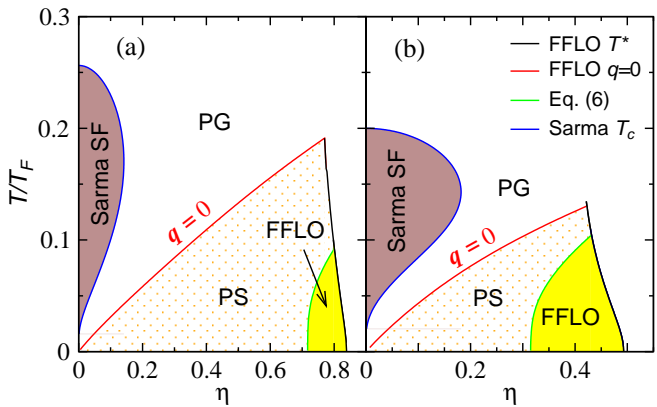}}
  \caption{ $T$--$\eta $ phase diagram of a homogeneous Fermi gas with
    equal mass at (a) $1/k_F^{}a =0$ and (b) $ -0.5$, corresponding to
    unitary and near-BCS cases, respectively. Here ``PG" and ``PS"
    indicate pseudogapped normal state and phase separation,
    respectively. An FFLO phase (yellow shaded) exists in the low $T$
    and relatively high $\eta $ regime, while they become unstable
    against phase separation in the dotted region. A beyond-mean-field
    Sarma superfluid lives in the intermediate $T$ and low $\eta $
    regime (brown shaded region). }
\label{fig:phasediagram}
\end{figure}

One may also compute the pseudogap, if well defined (as in a Sarma
phase), via
\begin{equation}
  \Delta_{pg}^{2}\equiv -\sum_P t(P) =a_0^{-1}\sum_{\textbf{p}}b({\Omega}_{\textbf{p}})  \,,
\label{eq:PG}
\end{equation}
where $b(x)$ is the Bose distribution function.
% and
%$\tilde{\Omega}_{\textbf{p}}=a_{0}\{\sqrt{1+4a_{1}
%  [B(\textbf{p}-\textbf{q})^{2}-\mu_{pair}]/a_{0}}-1\}/2a_{1}$ is the
%pair dispersion.
%
Just as the pairing fluctuations tend to destroy the condensate, the
presence of $\Delta_{pg}$ serves to deplete the order parameter
$\Delta_{sc}$ from the total excitation gap via $\Delta_{sc}^2 =
\Delta^2 -\Delta_{pg}^2$. In this case, pairing fluctuations will
reduce $T_c$ from its mean-field value to a lower temperature as
determined by $\Delta_{pg}=\Delta$.

For an equal-mass case, we take the majority (minority) species as
spin up (down) in our numerics.  For Fermi-Fermi mixtures, we take the
heavy (light) species to be spin up (down). In both cases, we take
Fermi momentum $k_{F}=(3\pi^{2}n)^{1/3}$, and define Fermi temperature
as $T_{F}=k_{F}^{2}/2m$, with $m=(m_{\uparrow}+m_{\downarrow})/2$.

We first present in Fig.~\ref{fig:phasediagram} the calculated
mean-field $T$--$\eta$ phase diagram for a homogeneous Fermi gas with
equal mass for the (a) unitary and (b) near-BCS cases,
respectively. Pairing takes place below the pairing temperature $T^*$
(black solid curve), where a pseudogap (PG) starts to emerge.  Here we
focus on the FFLO phase, not showing the boundary separating the
pseudogap state and the high $T$ normal phase, which is a crossover
rather than a true phase transition.  A mean-field FFLO state in the
low $T$ and relatively high $\eta$ regime for both cases. For lower
$\eta$, the FFLO states become unstable against phase separation (PS)
at low $T$ (dotted region), and these two phases are divided by the
green line, as determined by the stability condition
Eq.~(\ref{eq:sta}).  The red line denotes where $\mathbf{q}$ drops to
zero.  We also show the stable beyond-mean-field Sarma superfluid (SF)
phase (brown area) at intermediate $T$, as found previously
\cite{Chien06,LOFF1}. At the mean-field level, the PG phase would be
called the Sarma superfluid as well.

The phase diagram of the near-BEC case (say, $1/k_F^{}a=0.1$) is similar
but with a smaller phase space area for the FFLO states, which
eventually shrinks to zero towards the BEC regime. The counterpart
phase diagrams for Fermi-Fermi mixtures such as $^6$Li-$^{40}$K can be
found in Ref.~\cite{FFLO_MF_us}.

\begin{figure}
%\centerline{\includegraphics[clip,width=3.4in]{m1-inva-0.5-p0.45-T0.010-3D-full_bitmap.eps}}
%\centerline{\includegraphics[clip,width=3.4in]
%  {m1-inva0-p0.75-T0.010-alpha-p-omega-CC-full_2_bitmap.eps}}
\centerline{\includegraphics[clip,width=3.4in]
  {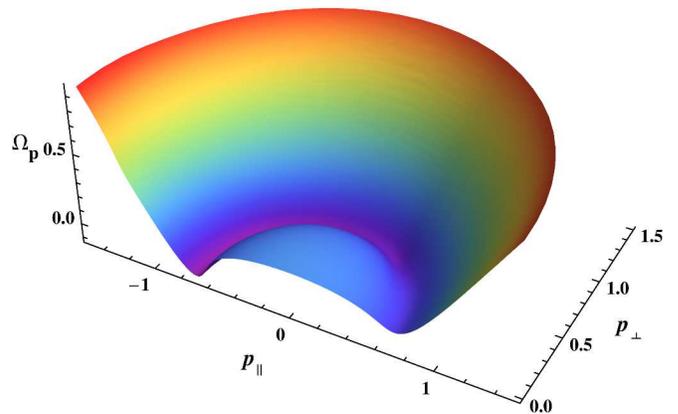}}
\caption{Typical pair dispersion $\Omega_\mathbf{p}$ in
  the FFLO phases in Fig.~\ref{fig:phasediagram}. Shown here is the
  unitary case with $\eta = 0.75$ and $T/T_F=0.01$.  The color coding
  is such that $\Omega_\mathbf{p}$ increases with the wavelength of
  the light. The units for energy and momentum are $E_F$ and $k_F$,
  respectively.}
\label{fig:3D}
\end{figure}

As a representative example, we next show in Fig. \ref{fig:3D} a 3D
plot of the pair dispersion $\Omega_\mathbf{p}$ in the FFLO phases. We
pick the unitary case in Fig.~\ref{fig:phasediagram}, with $\eta =
0.75$ and $T/T_F=0.01$, which has a solution of $q = 0.71$. Other
cases are very similar. The azimuthal angle $\theta$ in the plot
corresponds to the polar angle between $\mathbf{p}$ and $\mathbf{q}$
in the spherical coordinates in which we align $\mathbf{q}$ along the
$\hat{z}$ direction. It is evident that the rotational SO(3) symmetry
is broken.  And \emph{most importantly, the $\mathbf{p}=\mathbf{q}$
  point is a saddle point rather than the global minimum of the pair
  energy. }. An alternative plot at unitarity and counterpart plots
for the near-BCS and near BEC cases are shown in Supplementary
Figs.~S1 and S2, respectively.

\begin{figure}
\centerline{\includegraphics[clip,width=3.4in] {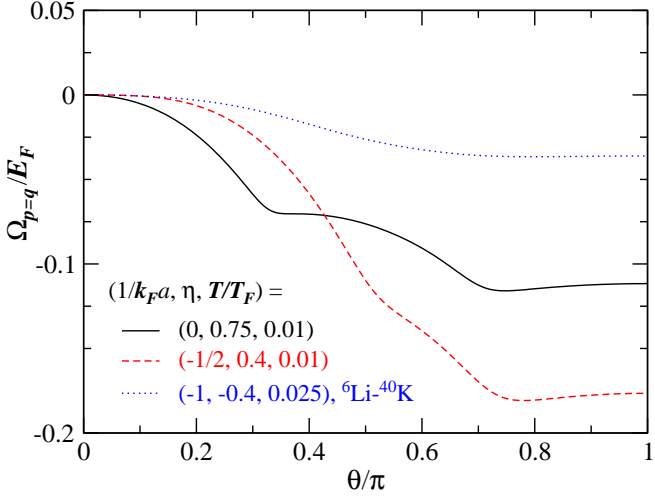}}
\caption{Pair energy $\Omega_{p=q}$ as a function of
  the polar angle $\theta$, i.e., along the bottom circle in the 3D
  plot shown in Fig.~\ref{fig:3D} (black solid line). Also shown are
  the near-BCS case (red dashed line) at $1/k_Fa =-1/2$ and
  $T/T_F=0.01$ with (equal mass and) population imbalance $\eta=0.4$,
  as well as the $^6$Li-$^{40}$K mixture case (blue dotted line) in
  the BCS regime with $1/k_Fa = -1$, $\eta = -0.4$ and
  $T/T_F=0.025$. For the latter case, the light species $^6$Li is the
  majority.}
\label{fig:minimum}
\end{figure}

To see this more clearly, we plot in Fig.~\ref{fig:minimum} the pair
dispersion $\Omega_{p=q}$ as a function of the polar angle $\theta$,
i.e., along the (constant radius) $p=q$ circle (in Fig.~\ref{fig:3D}),
as shown by the black solid line.  At the same time, we also show the
near-BCS case (red dashed line) at $1/k_Fa =-1/2$ with $\eta=0.4$ and
$T/T_F=0.01$, as well as the $^6$Li-$^{40}$K mixture case (blue dotted
line) in the BCS regime with $1/k_Fa = -1$, $\eta = -0.4$ and
$T/T_F=0.025$. In all cases, we find that the $\mathbf{p}=\mathbf{q}$
point is not the global minimum of the pair dispersion, in
contradiction to our assumption that the FFLO state is a spontaneously
broken symmetry state.  This means that the FFLO states found at the
mean-field level are \emph{not} stable once pairing fluctuations are
taken into account.

To make sure that this finding is not an artifact of the $G_0G$ scheme
of our $T$-matrix theory, we perform similar calculations using our
main competitor, the $GG$ scheme of the $T$-matrix theory, with
$\chi_{GG}^{}(P) =\sum_{K{\sigma}}
G_\sigma(P-K)G_{\bar{\sigma}}(K)/2$. This has been known as the FLEX
approximation \cite{FLEX1}, and have been used by various authors in
the study of BCS-BEC crossover. Then we compare the results between
these two schemes.

\begin{figure}
\centerline{\includegraphics[clip,width=3.3in] {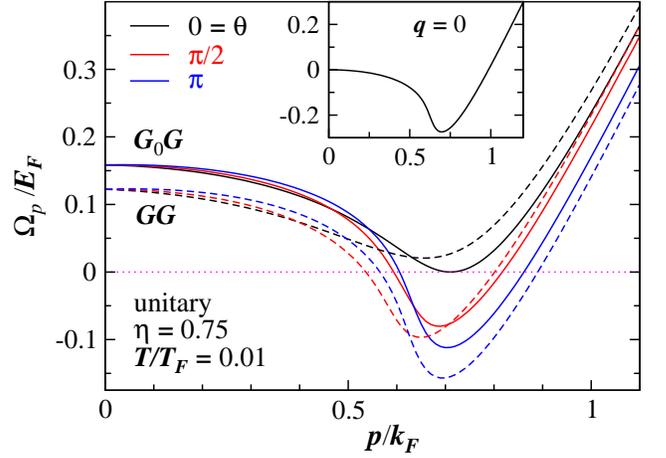}}
\caption{Pair dispersion $\Omega_{p}$ in the mean-field
  FFLO phase of a mass-balanced unitary Fermi gas with $\eta=0.75$ as
  a function of $p$ along different polar angles $\theta=0$ (black),
  $\pi/2$ (red), and $\pi$ (blue), for both the $G_0G$ (solid lines)
  and $GG$ (dashed lines) schemes of $T$-matrix theories. For neither
  scheme, the minimum energy along the $\mathbf{q}$ direction is the
  global minimum. Here the specific parameters are labeled. Shown in
  the inset is the pair dispersion with the same parameters but
  assuming a mean-field Sarma solution, for which $\mathbf{q}=0$.}
\label{fig:GG}
\end{figure}

Shown in Fig.~\ref{fig:GG} are representative pair dispersions
$\Omega_{p}$ as a function of $p$ along different polar angles
$\theta$ for both schemes, as labeled in the figure. To be specific,
we show the same unitary case as in Fig.~\ref{fig:3D}. The
corresponding 3D plot of the pair dispersion from the $GG$ scheme is
given in Supplementary Fig.~S3. For both schemes, we plot the curves
for $\theta=0$ (black), $\pi/2$ (red) and $\pi$ (blue).  Since the
$GG$ scheme (dashed lines) is inconsistent with the mean-field BCS gap
equation so that $U^{-1}+\chi^{}_{GG}(0,\mathbf{q}) \ne 0$, its pair
dispersion along $\theta=0$ does \emph{not} touch zero at its minimum,
unlike our $G_0G$ case (solid lines).
Nevertheless, common to both schemes is that the minimum pair energy
along $\theta=0$ is higher than along other directions, and is thus
\emph{not} a global minimum. 

By relaxing Eq.~(\ref{eq:min}), one may also study how a finite $q$
progressively leads to an angle dependence of the minimum of
$\Omega_p$, as shown in Supplementary Fig.~S4.

An unstable mean-field Sarma solution also exists in the mean-field
FFLO regime, with a typical pair dispersion shown in the inset of
Fig.~\ref{fig:GG} at high population imbalance.
Here $\Omega_{\mathbf{p}=0}$ vanishes, as determined by the gap
equation. However, the pair energy reaches a minimum at a finite $p$
on a 2D sphere $\mathbb{S}^2$ in the momentum space, which will
destroy the mean-field Sarma states. Most importantly, these pairs
will never Bose condense, and thus no symmetry breaking or phase
transition will occur. In this way, we have shown that the FFLO phase
will never occur in 3D continuum, as mentioned earlier.
Indeed, setting $\mu_{pair}$ to the bottom of the pair energy would
lead to a diverging noncondensed pair density, $n_{pair} =
a_0\Delta_{pg}^2$ via Eq.~(\ref{eq:PG}), and thus destroy
superfluidity.  Obviously, this divergence does not rely on the
rotational symmetry and can be readily extended to the case of optical
lattices. 

For a 2D case, there is no true long range order of superfluidity,
which is valid for zero momentum condensate. The dimensionality would
be reduced to 1D, leading to even stronger fluctuation effects, which
shall destroy FFLO type of superfluidity.

Similar instability of the FFLO states is also expected from the
$G_0G_0$ scheme of $T$-matrix approximation, as in the
Nozieres--Schmitt-Rink theory \cite{NSR}. It is easy to show that the
pair dispersion in the mean-field FFLO regime minimizes at a finite
$q$, as shown in Supplementary Fig.~S3. Indeed, using such a theory,
Ohashi also find that the FF state is unstable in 3D homogeneous Fermi
gases for a similar reason \cite{OhashiJPSJ71}.

While our calculations were done with the FF states, we argue that
such dimensional reduction effects hold for the LO and higher order
FFLO states as well. We shall also point out that the pairing field is
different from magnetic spin fluctuations, where unlike the pair
momentum, the magnitude of a spin is fixed so that spontaneous
symmetry breaking may occur as in a non-linear sigma model.

The reason FFLO states are unstable can be understood from a different
perspective. At the mean-field level, it is known that the LO states
has \emph{slightly} lower energy than the corresponding FF
states. While the FF states are condensation of Cooper pairs at a
single momentum $q$, the LO states are condensation at a pair of
momenta $\pm q$. It is conceivable that condensation at two pairs of
$q$'s forming a square in the momentum space shall further lower the
energy, as has been confirmed by mean-field calculations
\cite{ZhangShizhong}. Along the same line, it suggests that
condensation at 3, 4, 6, and 8 pairs of momenta and so on should have
a progressively lower energy. Eventually, it leads to the conclusion
that the lowest energy solution would be condensation on the entire 2D
constant-energy surface, on which the pair dispersion minimizes. This
is, of course, no longer a condensed state, nor an FFLO state. We note
that these mean-field crystalline states are different from ordinary
spontaneous breaking of the SO(3) symmetry, which typically has only
one preferential direction, as one finds in textbooks.

Finally, we investigate the nature of this unusual normal state, for
which the pair dispersion $\Omega_{p}$ minimizes at finite
$p$. The pairing correlation function for the 3D continuum case is
given by
\begin{eqnarray}
C(r) &\propto& \int \frac{\mbox{e}^{i\mathbf{p}\cdot
    \mathbf{r}}\mathrm{d}^3p}{\xi^2(p-q)^2+\tau }\nonumber\\&\approx&
\frac{1}{4\pi r\xi^2}\sqrt{\dfrac{4\xi^2q^2+\tau}{\tau}}\mbox{e}^{-r\sqrt{\tau}/\xi}
\sin(qr)\,,
\end{eqnarray}
where $\xi^2=a_0B_\parallel$ is the screening length (squared), and
$\tau = -a_0\mu_{pair} > 0$, with $\mu_{pair}\propto T$ near zero $T$. (Note
here that the pair dispersion is isotropic). Apart
from the oscillating behavior, the correlation length is given by
$\xi/\sqrt{\tau}\propto \xi/\sqrt{T}$. When $T\rightarrow 0$, the
exponential decay will disappear, leaving a $r^{-1}$ power law decay
at large distances so that the pairs approach an algebraic Bose
liquid. Note that this oscillating behavior due to a finite $q$ is
very unusual, manifesting the tendency to form a wave-like pairing
order. Without superfluidity, such a Bose liquid is a Bose metal in
the ground state, where $\mu_{pair}$ approaches 0 at zero $T$. Of course,
at high population imbalance, the major part of the system is composed
of the excessive majority fermions, which add to the metallic character
of the system. We shall call this phase ``anomalous metal''.

Recently, Radzihovsky and Vishwanath\cite{Ashvin} found that the LO
phase is unstable, which is consistent with our findings here. Further
on, they continued with the unstable LO state and concluded that
fermion pairs may pair again to form a nematic charge-4 SF$_4$
superfluid phase. However, because the interaction between fermion
pairs are usually repulsive, it is unlikely that such an SF$_4$ phase
will form. 

As of this writing, Boyack et al \cite{Rufus} found that the
superfluid density of a mean-field FF state vanishes in the direction
transverse to the wavevector $\mathbf{q}$, in agreement with our
findings here.

There have also been theoretical studies of possible FFLO (or stripe)
states in Fermi gases with spin-orbit coupling
\cite{chuanwei12,*xiaji13,Pu2013,*Iskin2013,*Yi2013PRL}. We point out
that the spin-orbit coupling forces a preferential direction, and/or
leads to topologically distinct Fermi surfaces, making the system
drastically different from the conventional FFLO physics. There are
also studies of FFLO phases in 1D Fermi gases, which, however, does
not process long range order at all.

In summary, we have studied the effects of pairing fluctuations on the
mean-field FFLO phases, and found that FFLO phases are intrinsically
unstable against pairing fluctuations in both continuum and optical
lattices in 3D and 2D, and thus do not exist experimentally. This
conclusion holds on general physical grounds, independent of our
specific pairing fluctuation theory, and is applicable for both
quantum gases and condensed matter systems.

We thank A.J. Leggett, K. Levin and S.Z. Zhang for helpful
discussions. This work is supported by NSF of China (Grant
No. 11274267), the National Basic Research Program of China (Grants
No. 2011CB921303 and No. 2012CB927404), NSF of Zhejiang Province of
China (Grant No.  LZ13A040001).

%\vspace*{-2ex}

%\bibliographystyle{apsrev} 
\bibliographystyle{naturemag} 
%\bibliographystyle{pnas-new} 
%\bibliography{Review3}

\begin{thebibliography}{10}
\expandafter\ifx\csname url\endcsname\relax
  \def\url#1{\texttt{#1}}\fi
\expandafter\ifx\csname urlprefix\endcsname\relax\def\urlprefix{URL }\fi
\providecommand{\bibinfo}[2]{#2}
\providecommand{\eprint}[2][]{\url{#2}}

\bibitem{FF}
\bibinfo{author}{Fulde, P.} \& \bibinfo{author}{Ferrell, R.~A.}
\newblock \bibinfo{title}{Superconductivity in a strong spin-exchange field}.
\newblock \emph{\bibinfo{journal}{Phys. Rev.}} \textbf{\bibinfo{volume}{135}},
  \bibinfo{pages}{A550--A563} (\bibinfo{year}{1964}).

\bibitem{LO}
\bibinfo{author}{Larkin, A.~I.} \& \bibinfo{author}{Ovchinnikov, Y.~N.}
\newblock \bibinfo{title}{Inhomogeneous state of superconductors}.
\newblock \emph{\bibinfo{journal}{Sov. Phys. JETP}}
  \textbf{\bibinfo{volume}{20}}, \bibinfo{pages}{762--769}
  (\bibinfo{year}{1965}).
\newblock \bibinfo{note}{[Zh. Eksp. Teor. Fiz. \textbf{47}, 1136 (1964)]}.

\bibitem{LOFF_Review}
\bibinfo{author}{Casalbuoni, R.} \& \bibinfo{author}{Nardulli, G.}
\newblock \bibinfo{title}{Inhomogeneous superconductivity in condensed matter
  and {QCD}}.
\newblock \emph{\bibinfo{journal}{Rev. Mod. Phys.}}
  \textbf{\bibinfo{volume}{76}}, \bibinfo{pages}{263--320}
  (\bibinfo{year}{2004}).

\bibitem{Radovan}
\bibinfo{author}{Radovan, H.~A.} \emph{et~al.}
\newblock \bibinfo{title}{Magnetic enhancement of superconductivity from
  electron spin domains}.
\newblock \emph{\bibinfo{journal}{Nature}} \textbf{\bibinfo{volume}{425}},
  \bibinfo{pages}{51--55} (\bibinfo{year}{2003}).

\bibitem{Kenzelmann}
\bibinfo{author}{Kenzelmann, M.} \emph{et~al.}
\newblock \bibinfo{title}{Coupled superconducting and magnetic order in
  {CeCoIn$_5$}}.
\newblock \emph{\bibinfo{journal}{Science}} \textbf{\bibinfo{volume}{321}},
  \bibinfo{pages}{1652--1654} (\bibinfo{year}{2008}).

\bibitem{ShimaharaJPSJ71}
\bibinfo{author}{Shimahara, H.}
\newblock \bibinfo{title}{Fulde-{F}errell-{L}arkin-{O}vchinnikov state and
  field-induced superconductivity in an organic superconductor}.
\newblock \emph{\bibinfo{journal}{J. Phys. Soc. Jpn.}}
  \textbf{\bibinfo{volume}{71}}, \bibinfo{pages}{1644} (\bibinfo{year}{2002}).

\bibitem{LebedPRB82}
\bibinfo{author}{Lebed, A.~G.} \& \bibinfo{author}{Wu, S.}
\newblock \bibinfo{title}{{Larkin-Ovchinnikov-Fulde-Ferrell} phase in the
  superconductor {(TMTSF)$_2$ClO$_4$}: Theory versus experiment}.
\newblock \emph{\bibinfo{journal}{\prb}} \textbf{\bibinfo{volume}{82}},
  \bibinfo{pages}{172504} (\bibinfo{year}{2010}).

\bibitem{GrafPRB72}
\bibinfo{author}{Vorontsov, A.~B.}, \bibinfo{author}{Sauls, J.~A.} \&
  \bibinfo{author}{Graf, M.~J.}
\newblock \bibinfo{title}{Phase diagram and spectroscopy of {FFLO} states of
  two-dimensional $d$-wave superconductors}.
\newblock \emph{\bibinfo{journal}{\prb}} \textbf{\bibinfo{volume}{72}},
  \bibinfo{pages}{184501} (\bibinfo{year}{2005}).

\bibitem{TingPRL96}
\bibinfo{author}{Wang, Q.}, \bibinfo{author}{Chen, H.-Y.}, \bibinfo{author}{Hu,
  C.-R.} \& \bibinfo{author}{Ting, C.~S.}
\newblock \bibinfo{title}{Local tunneling spectroscopy as signatures of the
  {F}ulde-{F}errell-{L}arkin-{O}vchinnikov state in $s$- and $d$-wave
  superconductors}.
\newblock \emph{\bibinfo{journal}{\prl}} \textbf{\bibinfo{volume}{96}},
  \bibinfo{pages}{117006} (\bibinfo{year}{2006}).

\bibitem{SimonsPRL102}
\bibinfo{author}{Berridge, A.~M.}, \bibinfo{author}{Green, A.~G.},
  \bibinfo{author}{Grigera, S.~A.} \& \bibinfo{author}{Simons, B.~D.}
\newblock \bibinfo{title}{Inhomogeneous magnetic phases: a {LOFF}-like phase in
  {Sr$_3$Ru$_2$O$_7$}}.
\newblock \emph{\bibinfo{journal}{\prl}} \textbf{\bibinfo{volume}{102}},
  \bibinfo{pages}{136404} (\bibinfo{year}{2009}).

\bibitem{ChoPRA83}
\bibinfo{author}{Cho, K.} \emph{et~al.}
\newblock \bibinfo{title}{Anisotropic upper critical field and a possible
  {F}ulde-{F}errel-{L}arkin-{O}vchinnikov state in a stoichiometric pnictide
  superconductor {LiFeAs}}.
\newblock \emph{\bibinfo{journal}{\prb}} \textbf{\bibinfo{volume}{83}},
  \bibinfo{pages}{060502R} (\bibinfo{year}{2011}).

\bibitem{Ptok2013}
\bibinfo{author}{Ptok, A.} \& \bibinfo{author}{Crivelli, D.}
\newblock \bibinfo{title}{The {Fulde-Ferrell-Larkin-Ovchinnikov state} in
  pnictides}.
\newblock \emph{\bibinfo{journal}{J. Low Temp. Phys.}}
  \textbf{\bibinfo{volume}{172}}, \bibinfo{pages}{226} (\bibinfo{year}{2013}).

\bibitem{SedrakianPRC67}
\bibinfo{author}{M\"uther, H.} \& \bibinfo{author}{Sedrakian, A.}
\newblock \bibinfo{title}{Phases of asymmetric nuclear matter with broken space
  symmetries}.
\newblock \emph{\bibinfo{journal}{\prc}} \textbf{\bibinfo{volume}{67}},
  \bibinfo{pages}{015802} (\bibinfo{year}{2003}).

\bibitem{Alford}
\bibinfo{author}{Alford, M.}, \bibinfo{author}{Bowers, J.} \&
  \bibinfo{author}{Rajagopal, K.}
\newblock \bibinfo{title}{Crystalline color superconductivity}.
\newblock \emph{\bibinfo{journal}{\prd}} \textbf{\bibinfo{volume}{63}},
  \bibinfo{pages}{074016} (\bibinfo{year}{2001}).

\bibitem{Bowers}
\bibinfo{author}{Bowers, J.~A.} \& \bibinfo{author}{Rajagopal, K.}
\newblock \bibinfo{title}{The crystallography of color superconductivity}.
\newblock \emph{\bibinfo{journal}{\prd}} \textbf{\bibinfo{volume}{66}},
  \bibinfo{pages}{065002} (\bibinfo{year}{2002}).

\bibitem{SedrakianPRA72}
\bibinfo{author}{Sedrakian, A.}, \bibinfo{author}{Mur-Petit, J.},
  \bibinfo{author}{Polls, A.} \& \bibinfo{author}{M\"uther, H.}
\newblock \bibinfo{title}{Pairing in a two-component ultracold {F}ermi gas:
  phases with broken space symmetries}.
\newblock \emph{\bibinfo{journal}{\pra}} \textbf{\bibinfo{volume}{72}},
  \bibinfo{pages}{013613} (\bibinfo{year}{2005}).

\bibitem{Liao}
\bibinfo{author}{Liao, Y.-A.} \emph{et~al.}
\newblock \bibinfo{title}{Spin-imbalance in a one-dimensional {Fermi} gas}.
\newblock \emph{\bibinfo{journal}{Nature}} \textbf{\bibinfo{volume}{467}},
  \bibinfo{pages}{567--569} (\bibinfo{year}{2010}).

\bibitem{Sheehy_RPP}
\bibinfo{author}{Radzihovsky, L.} \& \bibinfo{author}{Sheehy, D.~E.}
\newblock \bibinfo{title}{Imbalanced {F}eshbach-resonant {F}ermi gases}.
\newblock \emph{\bibinfo{journal}{Rep. Prog. Phys.}}
  \textbf{\bibinfo{volume}{73}}, \bibinfo{pages}{076501}
  (\bibinfo{year}{2010}).

\bibitem{Gloos}
\bibinfo{author}{Gloos, K.} \emph{et~al.}
\newblock \bibinfo{title}{Possible formation of a nonuniform superconducting
  state in the heavy-fermion compound {UPd$_2$Al$_3$}}.
\newblock \emph{\bibinfo{journal}{\prl}} \textbf{\bibinfo{volume}{70}},
  \bibinfo{pages}{501--504} (\bibinfo{year}{1993}).

\bibitem{Huxley}
\bibinfo{author}{Huxley, A.~D.} \emph{et~al.}
\newblock \bibinfo{title}{Flux pinning, specific heat and magnetic properties
  of the laves phase superconductor {CeRu}$_2$}.
\newblock \emph{\bibinfo{journal}{J. Phys. Condens. Matter}}
  \textbf{\bibinfo{volume}{5}}, \bibinfo{pages}{7709} (\bibinfo{year}{1993}).

\bibitem{NormanFFLOcomment}
\bibinfo{author}{Norman, M.~R.}
\newblock \bibinfo{title}{Existence of the {FFLO} state in superconducting
  {UPd$_2$Al$_3$}}.
\newblock \emph{\bibinfo{journal}{\prl}} \textbf{\bibinfo{volume}{71}},
  \bibinfo{pages}{3391} (\bibinfo{year}{1993}).

\bibitem{Tenya}
\bibinfo{author}{Tenya, K.} \emph{et~al.}
\newblock \bibinfo{title}{Field-history-dependent peak effect in the
  superconducting mixed state of {CeRu}$_2$}.
\newblock \emph{\bibinfo{journal}{Physica B: Condens. Matter}}
  \textbf{\bibinfo{volume}{259-261}}, \bibinfo{pages}{692--693}
  (\bibinfo{year}{1999}).

\bibitem{Review}
\bibinfo{author}{Chen, Q.~J.}, \bibinfo{author}{Stajic, J.},
  \bibinfo{author}{Tan, S.~N.} \& \bibinfo{author}{Levin, K.}
\newblock \bibinfo{title}{{BCS-BEC} crossover: From high temperature
  superconductors to ultracold superfluids}.
\newblock \emph{\bibinfo{journal}{Phys. Rep.}} \textbf{\bibinfo{volume}{412}},
  \bibinfo{pages}{1--88} (\bibinfo{year}{2005}).

\bibitem{Bloch_RMP}
\bibinfo{author}{Bloch, I.}, \bibinfo{author}{Dalibard, J.} \&
  \bibinfo{author}{Zwerger, W.}
\newblock \bibinfo{title}{Many-body physics with ultracold gases}.
\newblock \emph{\bibinfo{journal}{Rev. Mod. Phys.}}
  \textbf{\bibinfo{volume}{80}}, \bibinfo{pages}{885--964}
  (\bibinfo{year}{2008}).

\bibitem{LOFF1}
\bibinfo{author}{He, Y.}, \bibinfo{author}{Chien, C.-C.},
  \bibinfo{author}{Chen, Q.~J.} \& \bibinfo{author}{Levin, K.}
\newblock \bibinfo{title}{Single-plane-wave
  {L}arkin-{O}vchinnikov-{F}ulde-{F}errell state in {BCS-BEC} crossover}.
\newblock \emph{\bibinfo{journal}{Phys. Rev. A}} \textbf{\bibinfo{volume}{75}},
  \bibinfo{pages}{021602} (\bibinfo{year}{2007}).

\bibitem{SR06}
\bibinfo{author}{Sheehy, D.~E.} \& \bibinfo{author}{Radzihovsky, L.}
\newblock \bibinfo{title}{{BEC-BCS} crossover in ``magnetized"
  {F}eshbach-resonantly paired superfluids}.
\newblock \emph{\bibinfo{journal}{Phys. Rev. Lett.}}
  \textbf{\bibinfo{volume}{96}}, \bibinfo{pages}{060401}
  (\bibinfo{year}{2006}).

\bibitem{huhui06pra}
\bibinfo{author}{Hu, H.} \& \bibinfo{author}{Liu, X.-J.}
\newblock \bibinfo{title}{Mean-field phase diagrams of imbalanced {F}ermi gases
  near a {F}eshbach resonance}.
\newblock \emph{\bibinfo{journal}{Phys. Rev. A}} \textbf{\bibinfo{volume}{73}},
  \bibinfo{pages}{051603} (\bibinfo{year}{2006}).

\bibitem{helianyi06prb}
\bibinfo{author}{He, L.}, \bibinfo{author}{Jin, M.} \& \bibinfo{author}{Zhuang,
  P.}
\newblock \bibinfo{title}{Finite-temperature phase diagram of a two-component
  {F}ermi gas with density imbalance}.
\newblock \emph{\bibinfo{journal}{Phys. Rev. B}} \textbf{\bibinfo{volume}{74}},
  \bibinfo{pages}{214516} (\bibinfo{year}{2006}).

\bibitem{CombescotPRB7114}
\bibinfo{author}{Combescot, R.} \& \bibinfo{author}{Mora, C.}
\newblock \bibinfo{title}{Transition to the {Fulde-Ferrel-Larkin-O}vchinnikov
  planar phase : {A} quasiclassical investigation with {F}ourier expansion}.
\newblock \emph{\bibinfo{journal}{\prb}} \textbf{\bibinfo{volume}{71}},
  \bibinfo{pages}{144517} (\bibinfo{year}{2005}).

\bibitem{Yip07pra}
\bibinfo{author}{Yoshida, N.} \& \bibinfo{author}{Yip, S.-K.}
\newblock \bibinfo{title}{{Larkin-Ovchinnikov} state in resonant {F}ermi gas}.
\newblock \emph{\bibinfo{journal}{Phys. Rev. A}} \textbf{\bibinfo{volume}{75}},
  \bibinfo{pages}{063601} (\bibinfo{year}{2007}).

\bibitem{FFLO_MF_us}
\bibinfo{author}{Wang, J.~B.}, \bibinfo{author}{Che, Y.~M.},
  \bibinfo{author}{Zhang, L.~F.} \& \bibinfo{author}{Chen, Q.~J.}
\newblock \bibinfo{title}{Enhancement effect of mass imbalance on
  {F}ulde-{F}errell-{L}arkin-{O}vchinnikov type of pairing in {F}ermi-{F}ermi
  mixtures of ultracold quantum gases}.
\newblock \emph{\bibinfo{journal}{Sci. Rep.}} \textbf{\bibinfo{volume}{7}},
  \bibinfo{pages}{39783} (\bibinfo{year}{2017}).

\bibitem{Machida2}
\bibinfo{author}{Machida, K.}, \bibinfo{author}{Mizushima, T.} \&
  \bibinfo{author}{Ichioka, M.}
\newblock \bibinfo{title}{Generic phase diagram of fermion superfluids with
  population imbalance}.
\newblock \emph{\bibinfo{journal}{Phys. Rev. Lett.}}
  \textbf{\bibinfo{volume}{97}}, \bibinfo{pages}{120407}
  (\bibinfo{year}{2006}).

\bibitem{zw07pra}
\bibinfo{author}{Zhang, W.} \& \bibinfo{author}{Duan, L.-M.}
\newblock \bibinfo{title}{Finite-temperature phase diagram of trapped {F}ermi
  gases with population imbalance}.
\newblock \emph{\bibinfo{journal}{Phys. Rev. A}} \textbf{\bibinfo{volume}{76}},
  \bibinfo{pages}{042710} (\bibinfo{year}{2007}).

\bibitem{Kinnunen}
\bibinfo{author}{Kinnunen, J.}, \bibinfo{author}{Jensen, L.~M.} \&
  \bibinfo{author}{T\"orm\"a, P.}
\newblock \bibinfo{title}{Strongly interacting {F}ermi gases with density
  imbalance}.
\newblock \emph{\bibinfo{journal}{Phys. Rev. Lett.}}
  \textbf{\bibinfo{volume}{96}}, \bibinfo{pages}{110403}
  (\bibinfo{year}{2006}).

\bibitem{ZSSK06}
\bibinfo{author}{Zwierlein, M.~W.}, \bibinfo{author}{Schirotzek, A.},
  \bibinfo{author}{Schunck, C.~H.} \& \bibinfo{author}{Ketterle, W.}
\newblock \bibinfo{title}{Fermionic superfluidity with imbalanced spin
  populations}.
\newblock \emph{\bibinfo{journal}{Science}} \textbf{\bibinfo{volume}{311}},
  \bibinfo{pages}{492} (\bibinfo{year}{2006}).

\bibitem{Rice1}
\bibinfo{author}{Partridge, G.~B.}, \bibinfo{author}{Li, W.},
  \bibinfo{author}{Kamar, R.~I.}, \bibinfo{author}{Liao, Y.~A.} \&
  \bibinfo{author}{Hulet, R.~G.}
\newblock \bibinfo{title}{Pairing and phase separation in a polarized {F}ermi
  gas}.
\newblock \emph{\bibinfo{journal}{Science}} \textbf{\bibinfo{volume}{311}},
  \bibinfo{pages}{503} (\bibinfo{year}{2006}).

\bibitem{Stoof09prl}
\bibinfo{author}{Gubbels, K.~B.}, \bibinfo{author}{Baarsma, J.~E.} \&
  \bibinfo{author}{Stoof, H. T.~C.}
\newblock \bibinfo{title}{Lifshitz point in the phase diagram of resonantly
  interacting $^{6}\mathrm{Li}\mathrm{\text{-}}^{40}\mathbf{K}$ mixtures}.
\newblock \emph{\bibinfo{journal}{Phys. Rev. Lett.}}
  \textbf{\bibinfo{volume}{103}}, \bibinfo{pages}{195301}
  (\bibinfo{year}{2009}).

\bibitem{Stoof10pra}
\bibinfo{author}{Baarsma, J.~E.}, \bibinfo{author}{Gubbels, K.~B.} \&
  \bibinfo{author}{Stoof, H. T.~C.}
\newblock \bibinfo{title}{Population and mass imbalance in atomic {F}ermi
  gases}.
\newblock \emph{\bibinfo{journal}{Phys. Rev. A}} \textbf{\bibinfo{volume}{82}},
  \bibinfo{pages}{013624} (\bibinfo{year}{2010}).

\bibitem{Stoof12pra}
\bibinfo{author}{Baarsma, J.~E.} \& \bibinfo{author}{Stoof, H. T.~C.}
\newblock \bibinfo{title}{Inhomogeneous superfluid phases in
  $^{6}${Li}-$^{40}${K} mixtures at unitarity}.
\newblock \emph{\bibinfo{journal}{Phys. Rev. A}} \textbf{\bibinfo{volume}{87}},
  \bibinfo{pages}{063612} (\bibinfo{year}{2013}).

\bibitem{WuCJPRA83}
\bibinfo{author}{Cai, Z.}, \bibinfo{author}{Wang, Y.} \& \bibinfo{author}{Wu,
  C.}
\newblock \bibinfo{title}{Stable {Fulde-Ferrell-Larkin-O}vchinnikov pairing
  states in {2D} and {3D} optical lattices}.
\newblock \emph{\bibinfo{journal}{\pra}} \textbf{\bibinfo{volume}{83}},
  \bibinfo{pages}{063621} (\bibinfo{year}{2011}).

\bibitem{Buchleitner2012}
\bibinfo{author}{Franca, V.~V.}, \bibinfo{author}{H\"ordlein, D.} \&
  \bibinfo{author}{Buchleitner, A.}
\newblock \bibinfo{title}{Fulde-{F}errell-{L}arkin-{O}vchinnikov critical
  polarization in one-dimensional fermionic optical lattices}.
\newblock \emph{\bibinfo{journal}{Phys. Rev. A}} \textbf{\bibinfo{volume}{86}},
  \bibinfo{pages}{033622} (\bibinfo{year}{2012}).

\bibitem{Mendoza2013}
\bibinfo{author}{Mendoza, R.}, \bibinfo{author}{Fortes, M.},
  \bibinfo{author}{Sol\'is, M.~A.} \& \bibinfo{author}{Koinov, Z.}
\newblock \bibinfo{title}{Superfluidity of a spin-imbalanced {F}ermi gas in a
  three-dimensional optical lattice}.
\newblock \emph{\bibinfo{journal}{\pra}} \textbf{\bibinfo{volume}{88}},
  \bibinfo{pages}{033606} (\bibinfo{year}{2013}).

\bibitem{Koga2012}
\bibinfo{author}{Okawauchi, Y.} \& \bibinfo{author}{Koga, A.}
\newblock \bibinfo{title}{Stability of {FFLO} states in optical lattices with
  bilayer structure}.
\newblock \emph{\bibinfo{journal}{J. Phys. Soc. Jpn.}}
  \textbf{\bibinfo{volume}{81}}, \bibinfo{pages}{074001}
  (\bibinfo{year}{2012}).

\bibitem{Torma2012}
\bibinfo{author}{Kim, D.-H.} \& \bibinfo{author}{T\"orm\"a, P.}
\newblock \bibinfo{title}{Fulde-{F}errell--{L}arkin-{O}vchinnikov state in the
  dimensional crossover between one- and three-dimensional lattices}.
\newblock \emph{\bibinfo{journal}{\prb}} \textbf{\bibinfo{volume}{85}},
  \bibinfo{pages}{180508(R)} (\bibinfo{year}{2012}).

\bibitem{ChenAHai2012}
\bibinfo{author}{Chen, A.-H.} \& \bibinfo{author}{Gao, X.~L.}
\newblock \bibinfo{title}{Pure {Fulde-Ferrell-Larkin-O}vchinnikov state in
  optical lattices}.
\newblock \emph{\bibinfo{journal}{Phys. Rev. B}} \textbf{\bibinfo{volume}{85}},
  \bibinfo{pages}{134203} (\bibinfo{year}{2012}).

\bibitem{Chen2}
\bibinfo{author}{Chen, Q.~J.}, \bibinfo{author}{Kosztin, I.},
  \bibinfo{author}{Jank\'o, B.} \& \bibinfo{author}{Levin, K.}
\newblock \bibinfo{title}{Pairing fluctuation theory of superconducting
  properties in underdoped to overdoped cuprates.}
\newblock \emph{\bibinfo{journal}{Phys. Rev. Lett.}}
  \textbf{\bibinfo{volume}{81}}, \bibinfo{pages}{4708--11}
  (\bibinfo{year}{1998}).

\bibitem{footnoteonBECpairing}
\bibinfo{note}{When pairing is so strong that a two-body bound state with a
  large binding energy forms in the real space, the momenta of the component
  fermions in the pair will span a large momentum space. In this case, the
  {P}auli exclusion between the component fermions and the {F}ermi sphere is
  weak, and the pair will happily coexist with the Fermi sea, with a dispersion
  minimizing at zero momentum.}

\bibitem{ReviewLTP-Full}
\bibinfo{author}{Chen, Q.~J.}, \bibinfo{author}{Stajic, J.} \&
  \bibinfo{author}{Levin, K.}
\newblock \bibinfo{title}{Applying {BCS-BEC} crossover theory to high
  temperature superconductors and ultracold atomic {Fermi} gases}.
\newblock \emph{\bibinfo{journal}{Low Temp. Phys.}}
  \textbf{\bibinfo{volume}{32}}, \bibinfo{pages}{406} (\bibinfo{year}{2006}).
\newblock \bibinfo{note}{[Fiz. Nizk. Temp. \textbf{32}, 538 (2006)].}

\bibitem{OurRFReview}
\bibinfo{author}{Chen, Q.~J.}, \bibinfo{author}{He, Y.},
  \bibinfo{author}{Chien, C.-C.} \& \bibinfo{author}{Levin, K.}
\newblock \bibinfo{title}{Theory of radio frequency spectroscopy experiments in
  ultracold {F}ermi gases and their relation to photoemission experiments in
  the cuprates}.
\newblock \emph{\bibinfo{journal}{Rep. Prog. Phys.}}
  \textbf{\bibinfo{volume}{72}}, \bibinfo{pages}{122501}
  (\bibinfo{year}{2009}).

\bibitem{Chien06}
\bibinfo{author}{Chien, C.~C.}, \bibinfo{author}{Chen, Q.~J.},
  \bibinfo{author}{He, Y.} \& \bibinfo{author}{Levin, K.}
\newblock \bibinfo{title}{Intermediate temperature superfluidity in a {F}ermi
  gas with population imbalance}.
\newblock \emph{\bibinfo{journal}{Phys. Rev. Lett.}}
  \textbf{\bibinfo{volume}{97}}, \bibinfo{pages}{090402}
  (\bibinfo{year}{2006}).

\bibitem{chen07prb}
\bibinfo{author}{Chen, Q.~J.}, \bibinfo{author}{He, Y.},
  \bibinfo{author}{Chien, C.-C.} \& \bibinfo{author}{Levin, K.}
\newblock \bibinfo{title}{Theory of superfluids with population imbalance:
  Finite-temperature and {BCS-BEC} crossover effects}.
\newblock \emph{\bibinfo{journal}{Phys. Rev. B}} \textbf{\bibinfo{volume}{75}},
  \bibinfo{pages}{014521} (\bibinfo{year}{2007}).

\bibitem{Guo2009PRA}
\bibinfo{author}{Guo, H.}, \bibinfo{author}{Chien, C.-C.},
  \bibinfo{author}{Chen, Q.~J.}, \bibinfo{author}{He, Y.} \&
  \bibinfo{author}{Levin, K.}
\newblock \bibinfo{title}{Finite-temperature behavior of an interspecies
  fermionic superfluid with population imbalance}.
\newblock \emph{\bibinfo{journal}{Phys. Rev. A}} \textbf{\bibinfo{volume}{80}},
  \bibinfo{pages}{011601} (\bibinfo{year}{2009}).

\bibitem{wang13pra}
\bibinfo{author}{Wang, J.~B.}, \bibinfo{author}{Guo, H.} \&
  \bibinfo{author}{Chen, Q.~J.}
\newblock \bibinfo{title}{Exotic phase separation and phase diagrams of a
  {F}ermi-{F}ermi mixture in a trap at finite temperature}.
\newblock \emph{\bibinfo{journal}{Phys. Rev. A}} \textbf{\bibinfo{volume}{87}},
  \bibinfo{pages}{041601} (\bibinfo{year}{2013}).

\bibitem{PWY05}
\bibinfo{author}{Pao, C.-H.}, \bibinfo{author}{Wu, S.-T.} \&
  \bibinfo{author}{Yip, S.-K.}
\newblock \bibinfo{title}{Superfluid stability in the {BEC-BCS} crossover}.
\newblock \emph{\bibinfo{journal}{Phys. Rev. B}} \textbf{\bibinfo{volume}{73}},
  \bibinfo{pages}{132506} (\bibinfo{year}{2006}).

\bibitem{Stability}
\bibinfo{author}{Chen, Q.~J.}, \bibinfo{author}{He, Y.},
  \bibinfo{author}{Chien, C.-C.} \& \bibinfo{author}{Levin, K.}
\newblock \bibinfo{title}{Stability conditions and phase diagrams for
  two-component {F}ermi gases with population imbalance}.
\newblock \emph{\bibinfo{journal}{Phys. Rev. A}} \textbf{\bibinfo{volume}{74}},
  \bibinfo{pages}{063603} (\bibinfo{year}{2006}).

\bibitem{FLEX1}
\bibinfo{author}{Bickers, N.~E.}, \bibinfo{author}{Scalapino, D.~J.} \&
  \bibinfo{author}{White, S.~R.}
\newblock \bibinfo{title}{Conserving approximations for strongly correlated
  electron systems: {Bethe-S}alpeter equation and dynamics for the
  two-dimensional {H}ubbard model}.
\newblock \emph{\bibinfo{journal}{Phys. Rev. Lett.}}
  \textbf{\bibinfo{volume}{62}}, \bibinfo{pages}{961--964}
  (\bibinfo{year}{1989}).

\bibitem{NSR}
\bibinfo{author}{Nozi\`{e}res, P.} \& \bibinfo{author}{Schmitt-Rink, S.}
\newblock \bibinfo{title}{Bose condensation in an attractive fermion gas: from
  weak to strong coupling superconductivity}.
\newblock \emph{\bibinfo{journal}{J. Low Temp. Phys.}}
  \textbf{\bibinfo{volume}{59}}, \bibinfo{pages}{195--211}
  (\bibinfo{year}{1985}).

\bibitem{OhashiJPSJ71}
\bibinfo{author}{Ohashi, Y.}
\newblock \bibinfo{title}{On the {Fulde-F}errell state in spatially isotropic
  superconductors}.
\newblock \emph{\bibinfo{journal}{J. Phys. Soc. Jpn.}}
  \textbf{\bibinfo{volume}{71}}, \bibinfo{pages}{2625} (\bibinfo{year}{2002}).

\bibitem{ZhangShizhong}
\bibinfo{author}{Zhang, S.~Z.}
\newblock \bibinfo{note}{Private communications.}

\bibitem{Ashvin}
\bibinfo{author}{Radzihovsky, L.} \& \bibinfo{author}{Vishwanath, A.}
\newblock \bibinfo{title}{Quantum liquid crystals in an imbalanced {F}ermi gas:
  Fluctuations and fractional vortices in {L}arkin-{O}vchinnikov states}.
\newblock \emph{\bibinfo{journal}{\prl}} \textbf{\bibinfo{volume}{103}},
  \bibinfo{pages}{010404} (\bibinfo{year}{2009}).

\bibitem{Rufus}
\bibinfo{author}{Boyack, R.}, \bibinfo{author}{Wu, C.-T.},
  \bibinfo{author}{Anderson, B.~M.} \& \bibinfo{author}{Levin, K.}
\newblock \bibinfo{title}{Collective mode contributions to the {M}eissner
  effect: {F}ulde-{F}errell and pair-density wave superfluids}
  (\bibinfo{year}{2017}).
\newblock \bibinfo{note}{Unpublished}.

\bibitem{chuanwei12}
\bibinfo{author}{Zheng, Z.}, \bibinfo{author}{Gong, M.}, \bibinfo{author}{Zou,
  X.}, \bibinfo{author}{Zhang, C.} \& \bibinfo{author}{Guo, G.}
\newblock \bibinfo{title}{Route to observable
  {Fulde-Ferrell-Larkin-Ovchinnikov} phases in three-dimensional
  spin-orbit-coupled degenerate {F}ermi gases}.
\newblock \emph{\bibinfo{journal}{Phys. Rev. A}} \textbf{\bibinfo{volume}{87}},
  \bibinfo{pages}{031602} (\bibinfo{year}{2013}).

\bibitem{xiaji13}
\bibinfo{author}{Liu, X.-J.} \& \bibinfo{author}{Hu, H.}
\newblock \bibinfo{title}{Inhomogeneous {Fulde-Ferrell} superfluidity in
  spin-orbit-coupled atomic {F}ermi gases}.
\newblock \emph{\bibinfo{journal}{Phys. Rev. A}} \textbf{\bibinfo{volume}{87}},
  \bibinfo{pages}{051608(R)} (\bibinfo{year}{2013}).

\bibitem{Pu2013}
\bibinfo{author}{Dong, L.}, \bibinfo{author}{Jiang, L.} \& \bibinfo{author}{Pu,
  H.}
\newblock \bibinfo{title}{Fulde-{F}errell pairing instability in spin-orbit
  coupled {F}ermi gas}.
\newblock \emph{\bibinfo{journal}{New J. Phys.}} \textbf{\bibinfo{volume}{15}},
  \bibinfo{pages}{075014} (\bibinfo{year}{2013}).

\bibitem{Iskin2013}
\bibinfo{author}{Iskin, M.}
\newblock \bibinfo{title}{Spin-orbit coupling induced
  {Fulde-Ferrell-Larkin-O}vchinnikov-like {C}ooper pairing and skyrmion-like
  polarization textures in trapped optical lattices}.
\newblock \emph{\bibinfo{journal}{\pra}} \textbf{\bibinfo{volume}{88}},
  \bibinfo{pages}{013631} (\bibinfo{year}{2013}).

\bibitem{Yi2013PRL}
\bibinfo{author}{Wu, F.}, \bibinfo{author}{Guo, G.-C.}, \bibinfo{author}{Zhang,
  W.} \& \bibinfo{author}{Yi, W.}
\newblock \bibinfo{title}{Unconventional superfluid in a two-dimensional
  {F}ermi gas with anisotropic spin-orbit coupling and {Z}eeman fields}.
\newblock \emph{\bibinfo{journal}{\prl}} \textbf{\bibinfo{volume}{110}},
  \bibinfo{pages}{110401} (\bibinfo{year}{2013}).

\end{thebibliography}

\end{document}

%\setpagewiselinenumbers
%\modulolinenumbers[5]
%\linenumbers
%\setlength\columnsep{22pt}

\begin{center}{\Large Supplementary Information \\~ \\ Instability of
  Fulde-Ferrell-Larkin-Ovchinnikov states in three and two dimensions}

\vskip 1ex
{Jibiao Wang, Yanming Che, Leifeng Zhang and Qijin Chen$^*$}\\
\textit{Department of Physics and Zhejiang Institute of Modern
  Physics, Zhejiang University, Hangzhou, Zhejiang 310027, China and\\
 Synergetic Innovation Center of Quantum Information and
  Quantum Physics, Hefei, Anhui 230026, China} 
\end{center}

%\date{\today}

\vskip 2ex

Here we provide more data and plots, which serve as supplemental
information to the main text.
%\end{abstract}

%\pacs{03.75.Ss,03.75.Hh,67.85.-d,74.25.Dw}

%\maketitle

\vskip 4ex

%\section{Derivations of the pair dispersion for the GG scheme of T-matrix approximation }

\section{Pair dispersion in the mean-field FFLO phases from near-BCS through near-BEC regimes}

In this section, we will present more results of the pair dispersion at
high population imbalances in the \emph{mean-field} FFLO phase from
near-BCS through near-BEC regimes.

\begin{figure}
%\centerline{\includegraphics[clip,width=3.4in]
%  {m1-inva0-p0.75-T0.010-alpha-p-omega-CC-full_2_bitmap.eps}}
\centerline{\includegraphics[clip,width=4.in]
  {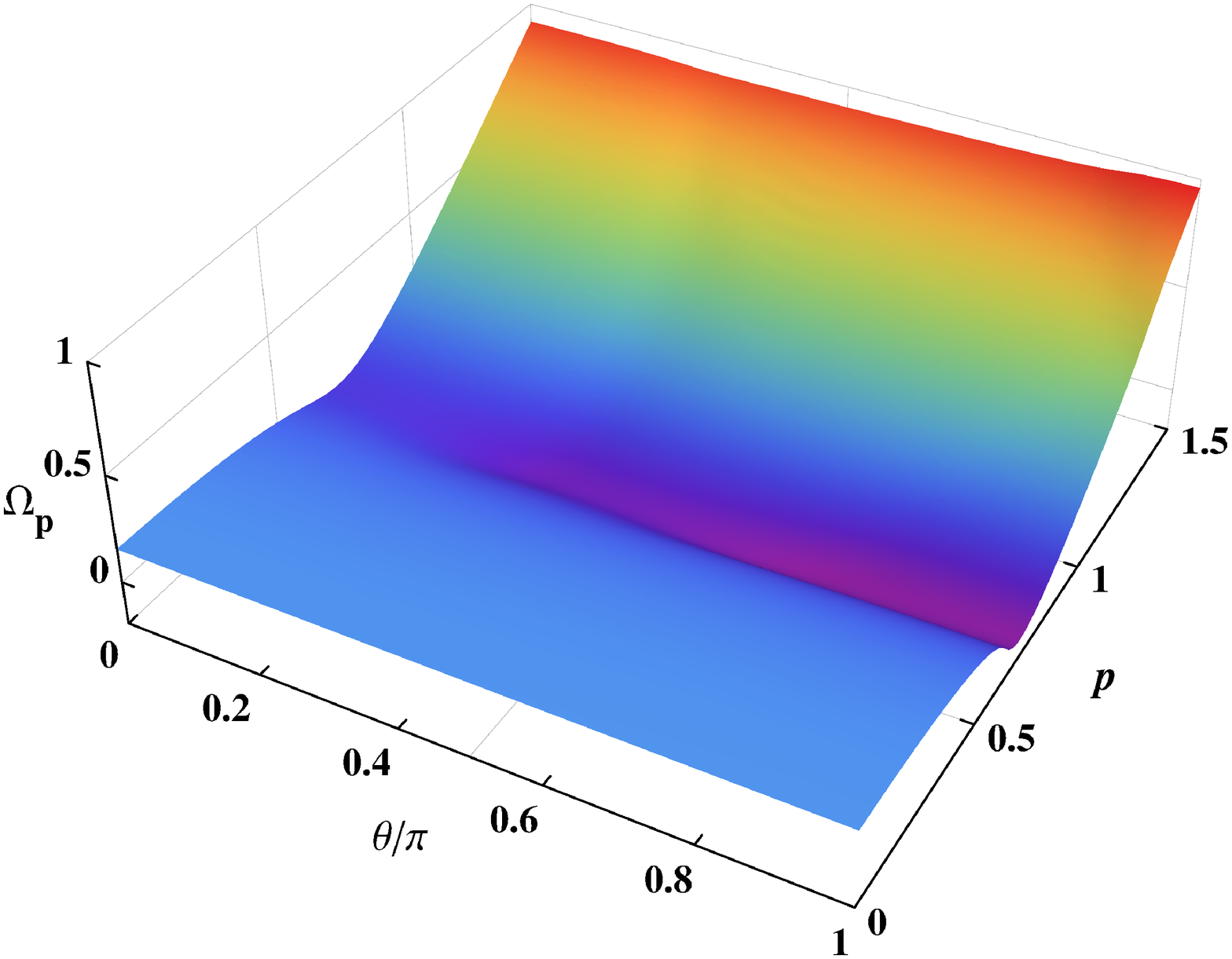}}
\caption{Alternative 3D plot of the pair dispersion
  $\Omega_\mathbf{p}$ in the FFLO phases at unitarity with population
  imbalance $\eta = 0.75$ and temperature $T/T_F=0.01$.  The
  conventions on color coding and units are the same as in
  Fig.~2 of the main text.}
\label{fig:3D_Unitary}
\end{figure}

Starting with the unitary case, Fig.~\ref{fig:3D_Unitary} shows the
pair dispersion $\Omega_\mathbf{p}$ in the FFLO phase with a population
imbalance $\eta = 0.75$ and equal masses at temperature
$T/T_F=0.01$. This is just an alternative 3D plot of Fig.~2 in the
main text, treating the angle $\theta$ between pair momentum
$\mathbf{p}$ and the FFLO wavevector $\mathbf{q}$ as a Descartes
coordinate. This makes it easier to see that the minimum value of
$\Omega_{p}$ (as a function of $p$) decreases as $\theta$ varies from
0 to $\pi$, revealing that the point $\mathbf{p=q}$ is indeed merely a
saddle point of $\Omega_\mathbf{p}$.

\begin{figure}
\centerline{\includegraphics[clip,width=3.2in]{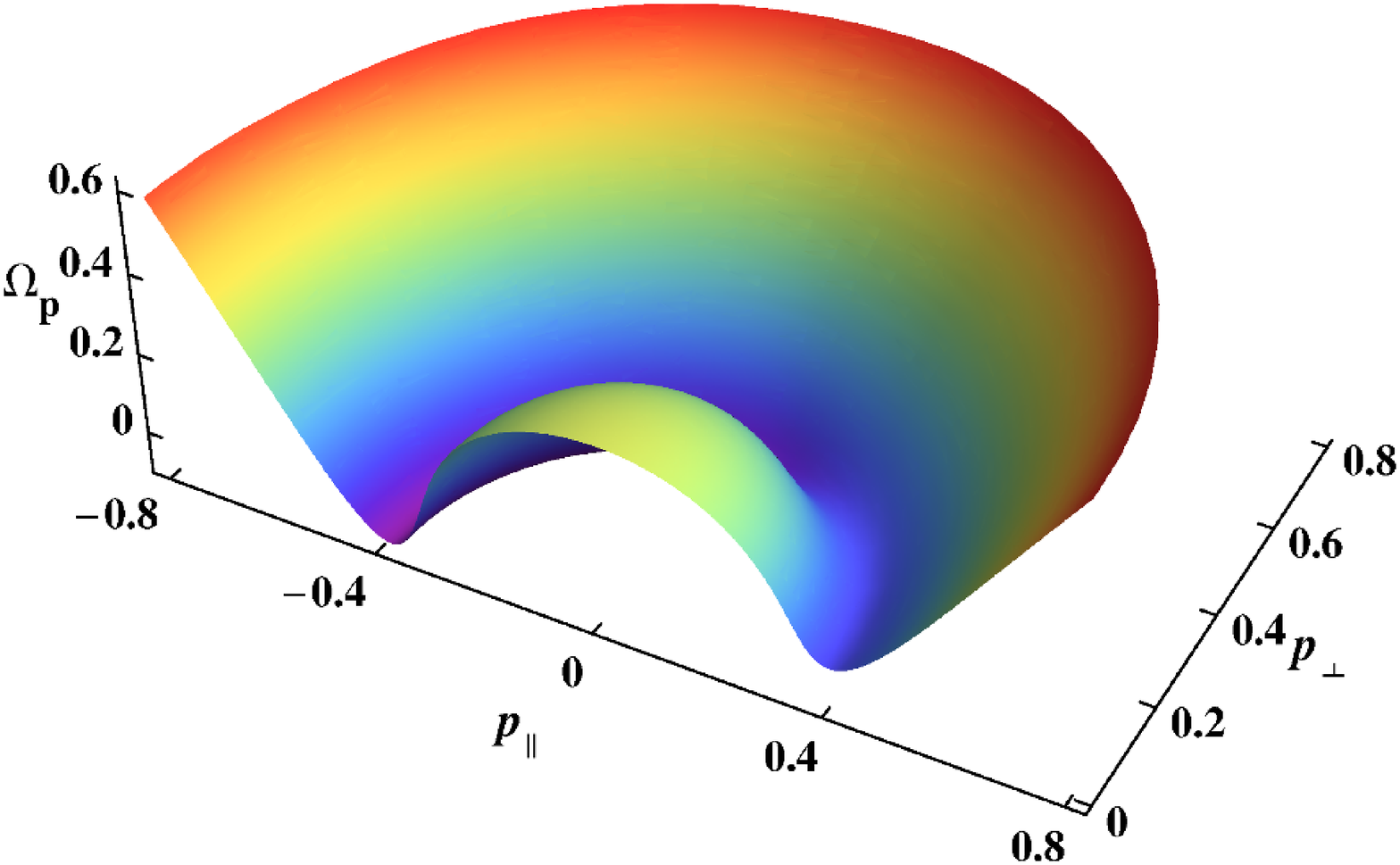}\hfill\includegraphics[clip,width=3.2in]{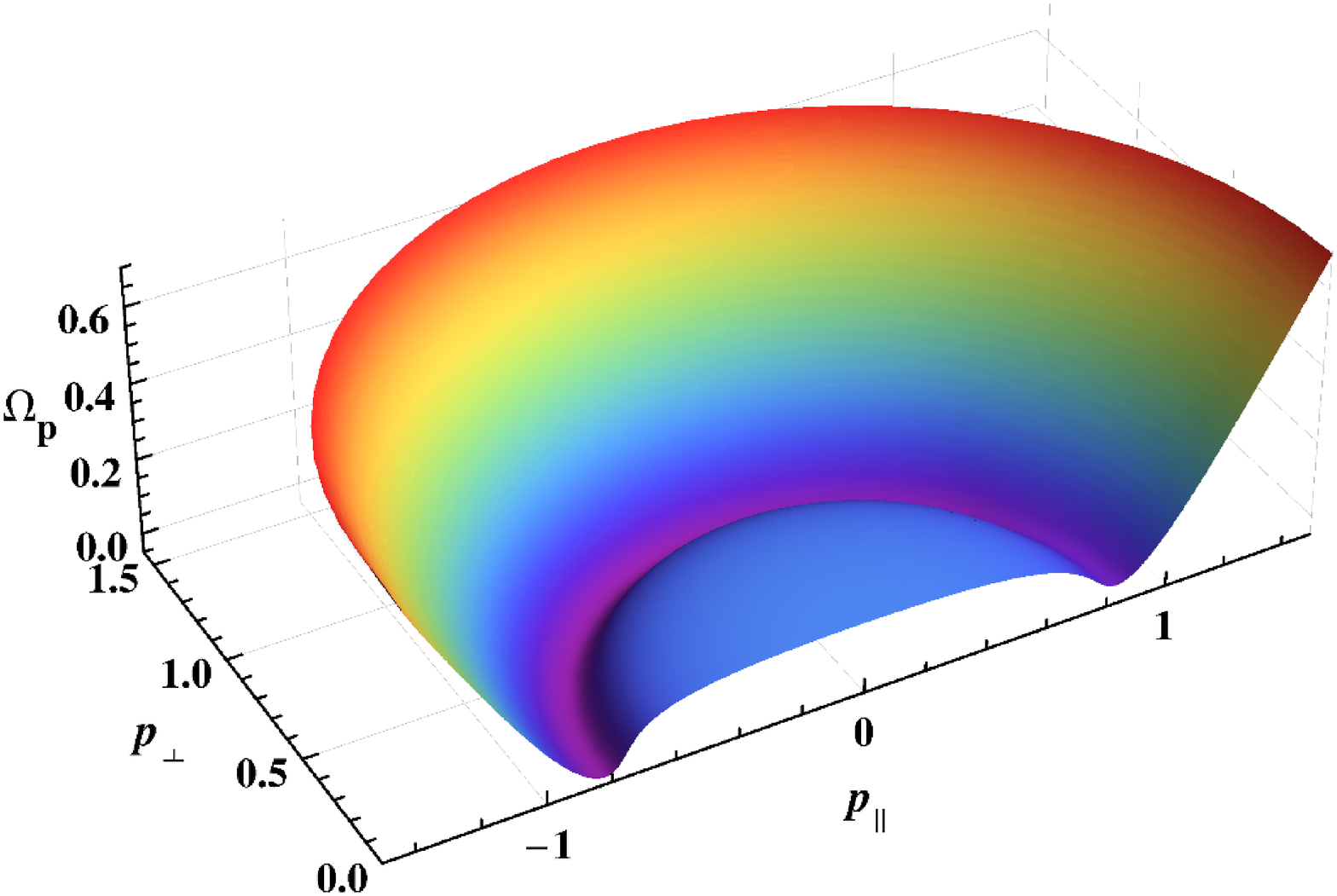}}
%\centerline{\includegraphics[clip,width=3.4in]
%{m1-inva0.1-p0.88-T0.010-3D_3.4in.eps}}
\caption{Pair dispersion $\Omega_\mathbf{p}$ in the FFLO phases for
  the near-BCS and near-BEC case, with $(1/k_Fa, \eta, T/T_F) = (-1/2,
  0.45, 0.01)$ and $(0.1, 0.75, 0.01)$ for the left and right panels,
  respectively. The conventions on color coding and units are the
  same as in Fig.~2 of the main text.}
\label{fig:3D_BCS_BEC}
\end{figure}

Now we show the counterpart plot of the near-BCS and near-BEC cases in
Fig.~\ref{fig:3D_BCS_BEC}, as the left and right panel,
respectively. Despite the different radii of the bottom (half) circle,
both confirms that the $\mathbf{p=q}$ point is a saddle point of
$\Omega_\mathbf{p}$.

%\begin{figure}
%\centerline{\includegraphics[clip,width=3.4in]{m1-inva-0.5-p0.45-T0.010-3D-half_3.4in.eps}}
%\centerline{\includegraphics[clip,width=3.4in]
%{m1-inva-0.5-p0.45-T0.010-3D_3.4in.eps}}
%\caption{Typical pair dispersion $\Omega_\mathbf{p}$ in the FFLO
%  phases in Fig.~1. Shown here is the near-BCS case with $1/k_Fa =
%  -1/2$, $\eta = 0.45$ and $T/T_F=0.01$.  The color coding is such
%  that $\Omega_\mathbf{p}$ increases with the wavelength of the
%  light. The units for energy and momentum are $E_F$ and $k_F$,
%  respectively.}
%\label{fig:3D_BCS}
%\end{figure}

\section{Pair dispersion from the $GG$ and $G_0G_0$ approximations of
  pairing fluctuation theories}

\begin{figure}
%\centerline{\includegraphics[clip,width=3.4in]
%  {GG_Invt_p-alpha_unitary.eps}}
%\centerline{\includegraphics[clip,width=3.4in]
%  {GG_Omega_theta_p_half_3.4in.eps}}
\centerline{\includegraphics[clip,width=3.2in]
  {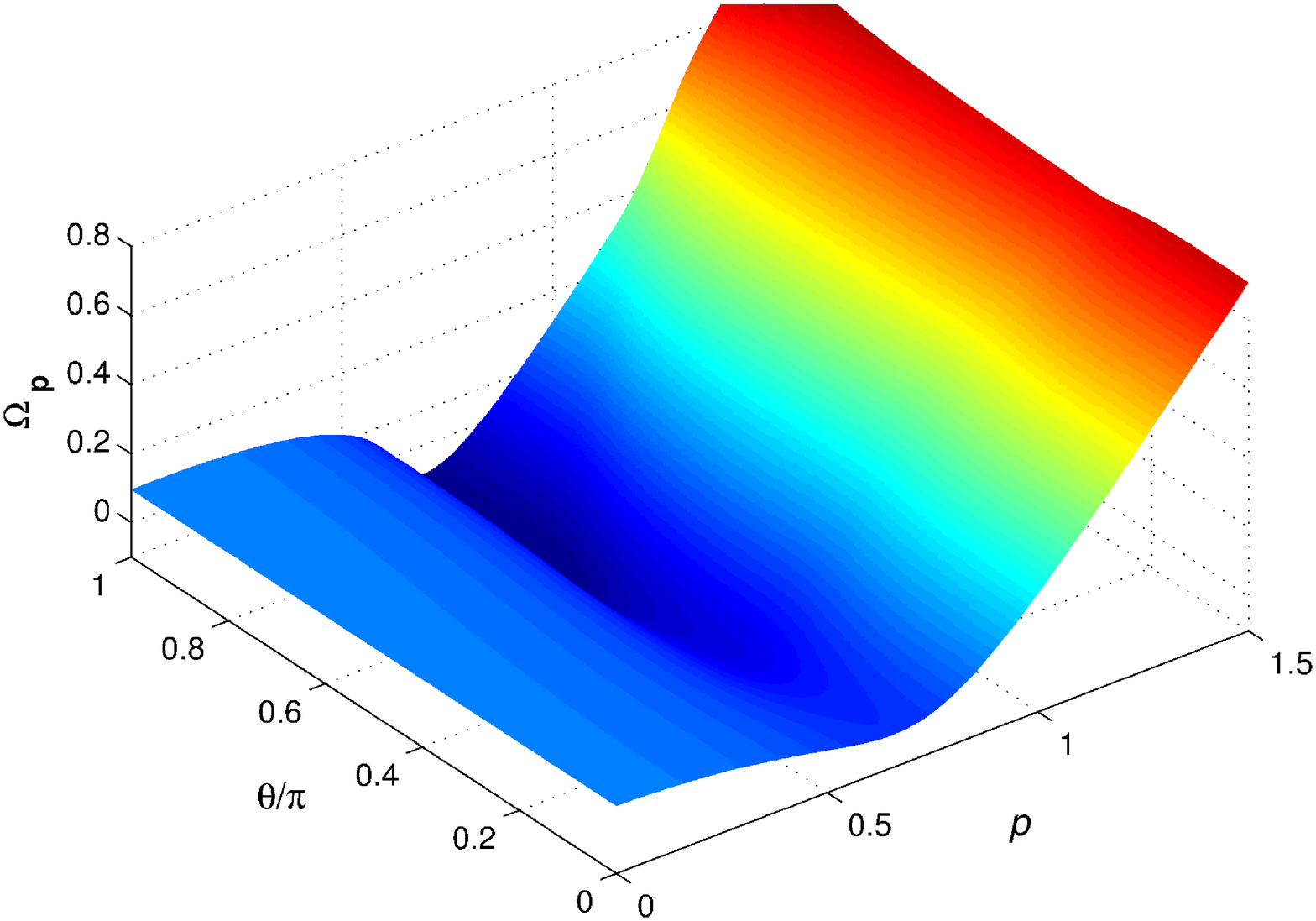}\hfill\includegraphics[clip,width=3.2in]
  {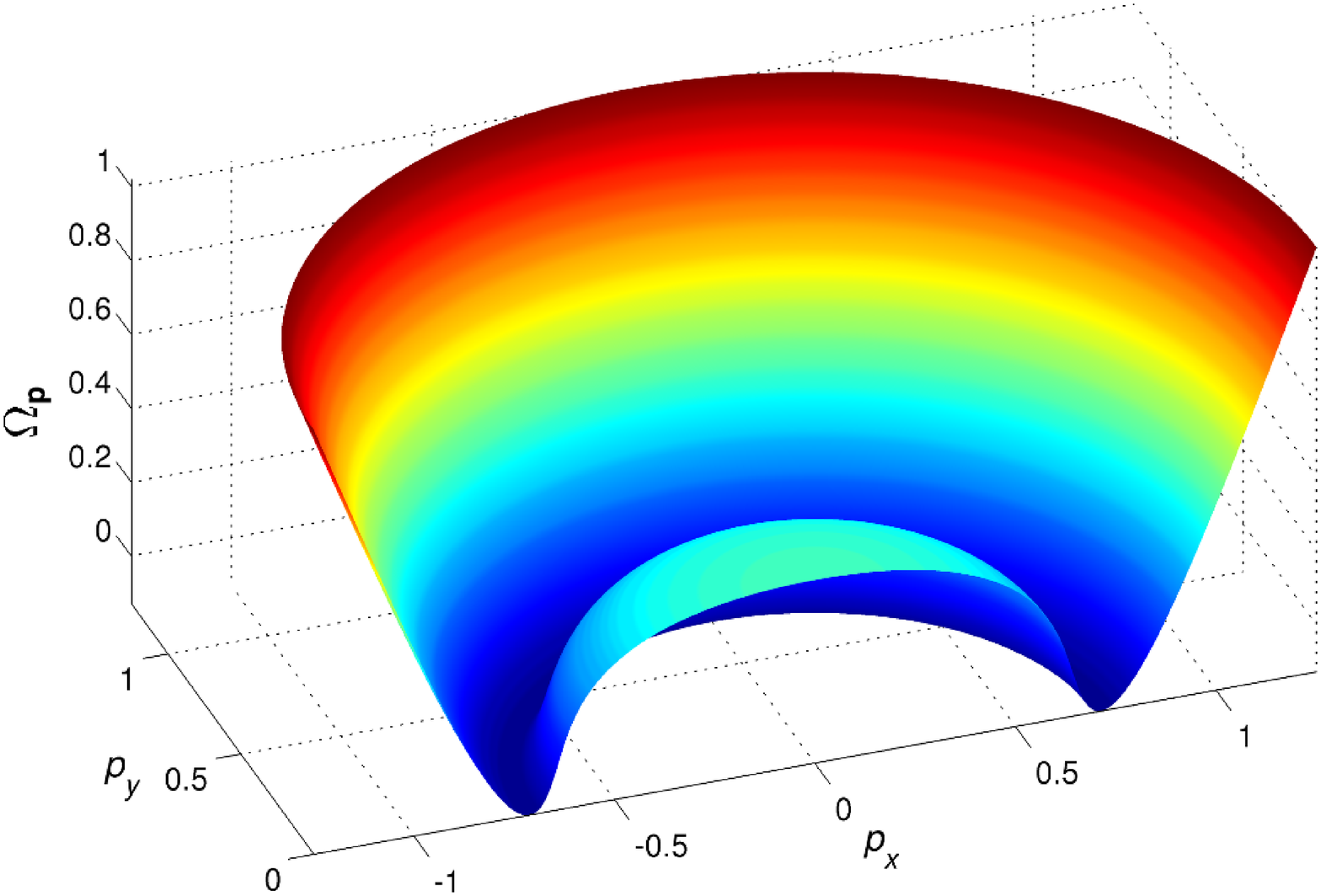}}
\caption{Typical pair dispersion $\Omega_\mathbf{p}$ in the mean-field
  FFLO phases for the (left) $GG$ and (right) $G_0G_0$ approximations
  of the pairing fluctuation theories. Shown here is the unitary case
  with $\eta = 0.75$ and $T/T_F=0.01$. The conventions on color coding
  and units are the same as in Fig.~2 of the main text.}
\label{fig:GG_NSR_unitary}
\end{figure}

Similar to the $G_0G$ scheme of the $T$-matrix approximation, one can
also extract the pair dispersion from the counterpart $T$ matrix in
the $GG$ and $G_0G_0$ schemes. The derivation is straight forward,
using by setting frequency $\Omega$ to zero in the inverse $T$ matrix,
as discussed above Eq. (7) in the main text. With no doubt, the
coefficient $a_0$ is quantitatively different. To make different plots
comparable in numerical values, we use the $a_0$ from the $G_0G$
scheme to plot the pair dispersion here. Note that one could
equivalently plot $-1/U + \chi(0,\mathbf{p})$ rather than
$\Omega_\mathbf{p}$ instead.  The resulting pair dispersion at
unitarity is shown in Fig.~\ref{fig:GG_NSR_unitary} for the (left)
$GG$ and (right) $G_0G_0$ schemes, respectively. Here values of the
chemical potentials $\mu_\sigma$, the gap $\Delta$, and the vector
$\mathbf{q}$ were the same as in Fig.~2 in the main text for the
$G_0G$ case.

Evidently, the pair dispersion for the $GG$ case is similar to that of
the $G_0G$ case, confirming that the $\mathbf{p=q}$ point is a saddle
point of $\Omega_\mathbf{p}$. In contrast, there is an obvious
difference between the $G_0G_0$ case and the other two; the pair
dispersion has no angle dependence. This can be easily understood
since the pair susceptibility $\chi_0(P) = \sum_K G_0(P-K)G_0(K)$ is
isotropic, independent of the gap $\Delta$ and the wavevector
$\mathbf{q}$. This is, of course, the defect of the
approximation. Nevertheless, this angle independence does suggest that
the pair energy minimizes on a finite momentum sphere so that no
spontaneous symmetry breaking or Bose condensation would take
place for the pairs.

Note that the pair dispersion for the $GG$ and $G_0G_0$ schemes does
not vanish at $\mathbf{p=q}$, because these two schemes are
incompatible with the BCS mean-field gap equation.

%\begin{figure}
%\centerline{\includegraphics[clip,width=3.4in]
%  {NSR_Omega_q_unitary.eps}}
%\centerline{\includegraphics[clip,width=3.4in]
%  {NSR_Omega_theta_p_half_3.4in.eps}}
%\caption{Typical pair dispersion $\Omega_\mathbf{p}$ in
%  the FFLO phases in . Shown here is the
%  unitary case with $\eta = 0.75$ and $T/T_F=0.01$.  The color coding
%  is such that $\Omega_\mathbf{p}$ increases with the wavelength of
%  the light. The units for energy and momentum are $E_F$ and $k_F$,
%  respectively.}
%\label{fig:NSR_unitary}
%\end{figure}

%\begin{figure}
%\centerline{\includegraphics[clip,width=3.4in]
%  {m1-inva-0.5-p0.45-p-G0G-GG_2.eps}}
%\caption{Typical pair dispersion $\Omega_\mathbf{p}$ in
%  the FFLO phases in . Shown here is the
%  unitary case with $\eta = 0.75$ and $T/T_F=0.01$.  The color coding
%  is such that $\Omega_\mathbf{p}$ increases with the wavelength of
%  the light. The units for energy and momentum are $E_F$ and $k_F$,
%  respectively.}
%\label{fig:Omegap-alpha_G0G+GG}
%\end{figure}

%\begin{figure}
%\centerline{\includegraphics[clip,width=3.4in]
%  {m1-inva-0.5-p0.45-T0.010-p-GG.eps}}
%\caption{Typical pair dispersion $\Omega_\mathbf{p}$ in
%  the FFLO phases in . Shown here is the
%  unitary case with $\eta = 0.75$ and $T/T_F=0.01$.  The color coding
%  is such that $\Omega_\mathbf{p}$ increases with the wavelength of
%  the light. The units for energy and momentum are $E_F$ and $k_F$,
%  respectively.}
%\label{fig:Omegap-p_GG}
%\end{figure}

\section{Evolution of pair dispersion with a forced LOFF wavevector $\mathbf{q}$}

\begin{figure}
\centerline{\includegraphics[clip,width=4.5in]
  {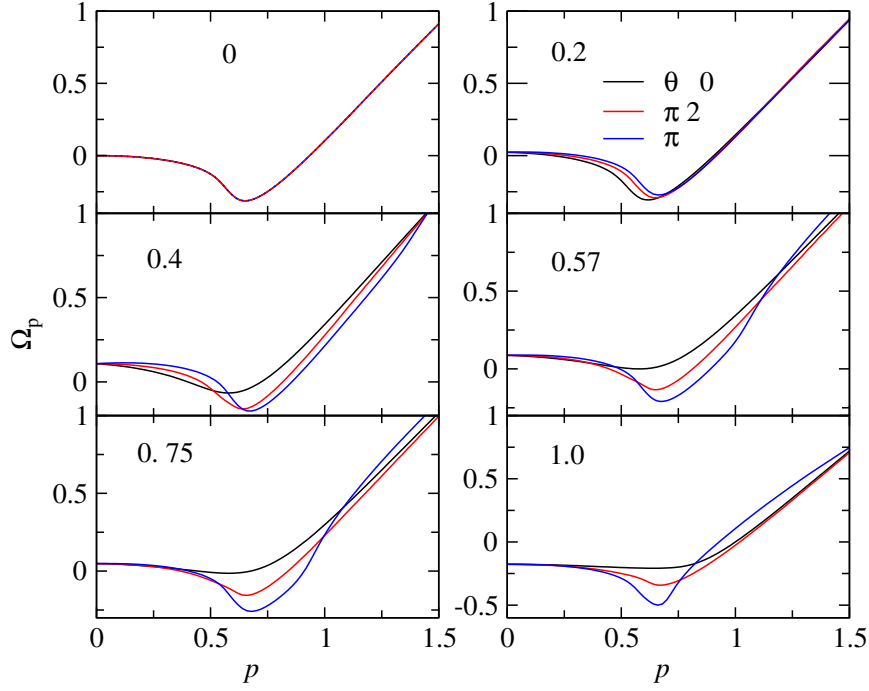}}
\caption{Evolution of the pair dispersion $\Omega_\mathbf{p}$ at
  different angles in the mean-field FFLO phases with increasing
  wavevector $q$, as labeled. Shown here is the unitary case with
  $\eta = 0.6$ at $T/T_F=0.01$.  }
\label{fig:Omegap-p_evolution}
\end{figure}

It is illuminating to show how the pairing dispersion evolves if one
forces and continuously tunes the FFLO wavevector $\mathbf{q}$,
starting from the Sarma solution in a mean-field FFLO phase. This can
be done by solving Eqs. (2)-(4) in the main text, without
Eq. (5). Here we work with our own pairing fluctuation theory, i.e.,
the $G_0G$ approximation. The result is shown in
Fig.~\ref{fig:Omegap-p_evolution} for $\eta = 0.6$ at unitarity. For
$q=0$, i.e., the Sarma state, the pair dispersion vanishes at zero
momentum, and is isotropic with no angle dependence.  The solution for
$q/k_F=0.578$ corresponds to the mean-field FF solution. As one can
see, angle dependence develops as $q$ increases from 0. In all finite
$q$ cases, the minimum pair energy along the $\mathbf{q}$ direction is
the highest among all angles. And therefore, these finite $q$ FFLO
states are unstable against pairing fluctuations.

%\section{Pair dispersion on an optical lattice}
%
%On a 3D optical lattice, the pair dispersion will minimize on a 2D constant
%energy surface. On a 2D optical lattice, it will minimize on a 1D
%constant energy circle. The latter case can be illustrated in
%Fig.~\ref{fig:Q2D}. In 
%
%\begin{figure}
%\centerline{\includegraphics[clip,width=3.4in]
%  {Omegaq2.eps}}
%\caption{Typical pair dispersion $\Omega_\mathbf{p}$ in
%  the FFLO phases in . Shown here is the
%  unitary case with $\eta = 0.75$ and $T/T_F=0.01$.  The color coding
%  is such that $\Omega_\mathbf{p}$ increases with the wavelength of
%  the light. The units for energy and momentum are $E_F$ and $k_F$,
%  respectively.}
%\label{fig:Q2D}
%\end{figure}

%\bibliographystyle{apsrev} 
%\bibliographystyle{naturemag} 
\bibliographystyle{pnas-new} 
%\bibliography{Review3}